\documentclass[12pt]{article}
\usepackage{graphicx}
\usepackage{mathptmx,mathrsfs}
\usepackage{amsfonts,amsmath,amssymb,amsthm}
\usepackage{indentfirst}
\usepackage{cite}
\usepackage{hyperref}
\usepackage{enumitem}
\usepackage{bm,graphics,color}
\usepackage{slashed}
\usepackage[english]{babel}

\numberwithin{equation}{section}
\theoremstyle{definition}

\begin{document}

\title{ Spontaneous symmetry breakings in  the singlet scalar  Yukawa model  within the
auxiliary
field method}

\author{A. A. Nogueira$^{1}$\thanks{andsogueira@hotmail.com}  \; and \;
F. L. Braghin$^{1}$\thanks{braghin@ufg.br}
\\
\textit{$^{1}$}\textit{\small{}Universidade Federal de Goi\' as, Instituto de F\'isica, 
Av. Esperan\c{c}a, 74690-900, Goi\^ania, Goi\' as, Brasil}\\
}

\maketitle
\date{}

\begin{abstract}

The aim of this work is to investigate the occurrence of two different
spontaneous symmetry breakings {at} two levels of the description of fermion-scalar field 
model, by means of a set of gap equations and {with} a background field effective action.
For that, we consider the Yukawa model, as a toy
model for interactions between non-massive fermions intermediated by a 
 self-interacting real  scalar field.
 This model has at stakes two symmetries at the classical level that, as we know, 
 might be spontaneously or dynamically broken  with  mass generation for the particles. 
The   auxiliary field method is considered 
and it  produces
 coupled renormalized gap  equations.
The effective action is then written  with  quantum contributions by {the} external background {field} method.
 We brought to light   how the renormalization procedure affects the physical gaps, investigate its properties, and discuss the connection between the auxiliary fields not only to define composite states but also to compute the effective action.
 
\end{abstract}
\newpage{}

\section{Introduction}

A well known  method to deal with polynomial self interactions in field theories is
the auxiliary field method (AFM), 
or {the} so called Hubbard-Stratonovich (HS) {transformations} \cite{HubbStra,Bozoni},
that 
provides an interesting way {to deal} with {the} non linear dynamics.
This method  is  also suitable to investigate spontaneous or dynamical symmetry breakings
(SSB or DSB)
that have important roles {inside} the Standard Model (SM)  \cite{SM-SB,ChiConf}. 
The auxiliary field method   may have a particular extra advantadge that is to define {the}
composite states
 that might correspond to {quasi-particles} of the system,
as it is the case for the definition of mesons states in 
Strong Interactions \cite{NJL1,Klev,Ebert}.
This method  might  be extended for higher order interactions
in different ways
\cite{coimbra,alkofer,FLB-epjp}.
Besides that, 
 observables in a quantum field theory
 might be described in terms of
vacuum expected values of composite fields
that can give  rise to  different condensates
 \cite{Weinberg,Mosel}.
In the present work we intend to revisit some 
of these issues.
For that,
we will consider 
the   Yukawa model (YuM)
with a self interacting scalar singlet
  wherein we consider massless 
fermions whose masses are generated by  SSB 
\cite{YuMo1}.
This model can be seen  as a prototype to understand
the Higgs sector of the SM  \cite{YuMo1,Frasca}
and, for example, to address the role of discretization of fermions in lattices
\cite{lattice-1,lattice-2}.
In this program, the spontaneous symmetry breaking of the  $Z_{2}$ discrete symmetry 
 takes place \cite{peskin}.
Besides the interest in considering toy models, such as the scalar singlet {Higgs-Yukawa},
to develop analytical techniques and to have a path  to analyze phenomenology,
extensions of the Standard Model with an additional singlet Higgs 
have been envisaged \cite{ext-SM}.
As a model{,} it might eventually  reproduce aspects of the fundamental {and} more intrincated  theory.
It has also been considered, for example,   for the analysis of a scalar field 
 with some modifications \cite{Gies-etal}, in an external gravitational
 field \cite{Shap}, analyzed by means of holography \cite{Wu-etal},
envisaged for dark matter investigations \cite{Herrero-etal} 
and its phase diagram has been addressed extensively  in the large $N_f$ limit
   \cite{largeNf,rosenstein-etal}.
In spite  of the  different approaches employed to understand further 
the YuM and its scalar sector{,}
 one can rather reach {also} upper and lower bounds
for  masses and coupling constants \cite{bock-etal,gerhold-jansen}.
Its  renormalization has been very often employed at one loop level
by means of 
the effective action technique introduced by 
Coleman-Weinberg to describe the origin of spontaneous symmetry breaking \cite{Coleman}.
One can expect that the  YuM, in particular its ground state, eventually  {can} be  suitably described 
by a series of condensates, $<\phi>,   <\phi^2>,   <\bar{\psi}\psi>$ and so on.
These three condensates, by the way, can be considered 
 the leading lower dimensional ones.
The other one would be $<\phi^3>$ that will be considered
 to  be factorized into the 
first two of them.

The exchange of scalar field in the { Yukawa model} provides a mechanism for fermion interaction 
that cannot be currently tested, although  power counting arguments can also lead 
 to quark contact interactions at the energies scales
of LHC \cite{contact-lhc}. 
The Higgs particle  might participate {into} this type of mechanism for heavier quarks.
The investigation of the role of the quartic vertices
 and their relationship with symmetry and mass has a relevant role in physics{,} corresponding to an effective description of the 
interactions that eventually should find justification in {a} more fundamental
  boson mediation processes
explored in 
many other contexts \cite{Fermi,Nambu,Chiral,Sigmamodel,Renoron,GrossNev}.
The relation of fundamental (renormalizable) theories and {the} effective models 
has been explored in the last decades and it helped the construction of the SM
with its interplay with phenomenology and therefore it  allowed the discovery of many 
effects and phenomena.
Effective models 
are expected to be valid in a restricted range 
of energies, usually low energies with respect to an energy/momentum scale 
or  {cut-off} $\Lambda$.
Usually they are non-renormalizable.
The model  may be renormalized at a particular level of calculation
and for each new quantum correction there might arise
the  need of systematic changes or corrections  
   \cite{eff2,eff,Weinberg,Predazzi,Renorm1,Renorm2,NJL-renorm}.
Although the SSB effect keeps some different characteristics
from Dynamical Chiral Symmetry Breaking (DChSB),
they have several properties in common { that might be implemented }
not only in the full version of the SM but 
also in effective models
\cite{Polchi}.
The occurrence of SSB or dynamical symmetry breaking is usually directly related to
 the phenomenology of  gap equations, widely present in effective models
for the strong and weak interactions,
 with the  implicit 
 discussion of the mechanisms for {mass generation}.
Historically this has been initially  envisaged  in the scalar case  \cite{Salam}
 and also in the vector case
with its connection with the gauge symmetry 
\cite{Shaw,Deser}. 
Later the renormalization of the scalar or gauge theories with 
and without symmetry breaking
have been established \cite{R1,R2,R3,R4,R5,R6}.

In this work, 
we investigate
the   Yukawa interaction between 
massless fermions and a self interacting scalar
by means of the 
auxiliary field technique.
When reducing the original model by {the} auxiliary composite fields{,}
both mechanisms of spontaneous and dynamical symmetry breakings,
providing mass generations,
can be explored independently or simultaneously, depending on the values of the masses and coupling constants.
As a second level of analysis,  we address the renormalization aspect of the 
resulting auxiliary field effective action,  using the logics of
 \cite{Paulo-Braghin} for a different model.
We show the consequences of the gap renormalization,
 unveiling the properties of the renormalized fermion-boson system,
for a single component scalar field.
We find out the link between the auxiliary fields and the vertices in the effective action by the current expansion methodology. 
The paper is organized as follows: 
In Sec. \ref{Sec1} we 
derive  the coupled renormalized gap  equations:  
one for the YuM composite scalar 
field, $\Psi$, and the other for 
a  composite fermion condensate $<\bar{\psi} \psi>$ as a single flavor 
chiral condensate. We investigate too the proprieties of the coupled  gap equations for particular 
limits 
 of $m_{R}$, $\lambda_{R}$ and $g_{R}$. In Sec. \ref{Sec12} we write the effective action from quantum contributions 
considering  background field   
methods and study the consequence of renormalization in the masses and coupling constants. 
We explore also the conditions for the existence of two-boson state and fermion-antifermion state.
 For that, mixing interaction between the composite fields
describing the two-boson and two-fermion states is identified.
In Sec. \ref{Sec3} there are final remarks.

\section{Composite fields and the coupled  gap  equations}
\label{Sec1}

By starting
with  the YuM with SSB one reaches  a non-renormalizable  effective model 
that includes {the} Fermi-type {of} fermion interactions, with 
the contribution of  a boson condensate {  $\phi_0 = < \phi >_0$.
The subscript $_0$ indicates a condensate valued in the vacuum.
 In the limit 
 of a  resulting 
 effective four-fermion interaction
reasonably strong, one can also obtain a 
DChSB.

\subsection{ Composite-scalar field }
\label{Sec11}

The generating functional of 
 {the} Yukawa model for  massless fermions coupled to massive self interacting 
scalar field can be written  as:
\begin{eqnarray}
Z &=& N \int {\cal D}[\phi, \bar{q}, q]\exp[i\int d^{4}x \; ( {\cal L} + {\cal L}_{s} ),\cr\cr
{\cal L}&=&\bar{q}(i\slashed{\partial}-g\phi)q+\frac{1}{2}(\partial_{\mu}\phi)^{2}+m^{2}\phi^{2}-\frac{\lambda}{4!}\phi^{4},
\end{eqnarray}
{wherein} the scalar and fermion field sources ($J$ and $\eta, \bar{\eta}$) are encoded in the term:
${\cal L}_s =   ( J \phi + \bar{\eta} q + \bar{q} \eta )$.
The scalar field sector is subject to the spontaneous symmetry breaking
depending on the scalar field mass, therefore at this tree level the usual 
conditions for the emergence of the so-called  scalar field condensate {is the following}:
\begin{eqnarray}  \label{ssb-tree}
 \bar\phi^2_{0}  = \frac{12 m^2}{\lambda},
\end{eqnarray}
in} which one needs $m^2 > 0$. {These conditions will receive corrections due to the quantization of the scalar and of the fermion fields and it will be discussed again below.}

With the renormalization procedure we establish the following relation between the naked and dressed fields and parameters:
\begin{eqnarray}
&&\phi={\cal Z}_{\phi}^{\frac{1}{2}}\phi_{R},\quad q={\cal Z}_{q}^{\frac{1}{2}}q_{R},\quad {\bar q}={\cal Z}_{q}^{\frac{1}{2}}{\bar q}_{R},\cr\cr
&&m=\frac{{\cal Z}_{m}}{{\cal Z}_{\phi}}m_{R},\quad \lambda=
\frac{{\cal Z}_{\lambda}}{{\cal Z}_{\phi}^{2}}\lambda_{R}, \quad g=\frac{{\cal Z}_{g}}{{\cal Z}_{q}{\cal Z}_{\phi}^{\frac{1}{2}}}g_{R}.
\end{eqnarray}
With these redefinitions the model is then be written by:
\begin{equation}
\label{Lagrern}
{\cal L}={\cal Z}_{q}\bar{q}_{R}i\slashed{\partial}q_{R}-{\cal Z}_{g}g_{R}\phi_{R}\bar{q}_{R}q_{R}+{\cal Z}_{\phi}\frac{1}{2}(\partial_{\mu}\phi_{R})^{2}+{\cal Z}_{m}m_{R}^{2}\phi_{R}^{2}-{\cal Z}_{\lambda}\frac{\lambda_{R}}{4!}\phi_{R}^{4}.
\end{equation}
wherein we see a linear combination of all terms of the original Lagrangian respecting the discrete $Z_{2}$ symmetry and global charge conservation.
In terms of 
 the counter-terms notation ${\cal Z}_{i}=1+\delta{\cal Z}_{i}$ we write the previous equation as 
\begin{eqnarray}
\label{Lr}
&&{\cal L}={\cal L}_{R}^{free}+{\cal L}_{R}^{c.t},\cr\cr
&&{\cal L}_{R}^{c.t}=\delta{\cal Z}_{q}\bar{q}_{R}i\slashed{\partial}q_{R}-
\delta {\cal Z}_{g}g_{R}\phi_{R}\bar{q}_{R}q_{R}+\delta{\cal Z}_{\phi}\frac{1}{2}(\partial_{\mu}\phi_{R})^{2}+\delta{\cal Z}_{m}m_{R}^{2}\phi_{R}^{2} - 
\delta{\cal Z}_{\lambda}\frac{\lambda_{R}}{4!}\phi_{R}^{4}.
\end{eqnarray}
 From now on, we will only present the renormalization 
of some of the resulting  equations in which a more complete discussion can be done.

With the background or external field method
the scalar field can be shifted by a background part
 as $\phi\rightarrow\phi_{0}+\tilde{\phi}$ where
 $\phi_{0}$ is the background part.
In a first analysis  $\phi_0$
is  the classical homogeneous condensate $\bar{\phi}_{0}$, although
 we let the freedom for receiving corrections latter. 
The field $\tilde{\phi}$
will   be integrated out.
To make possible quantization of this scalar field  
the AFM is used by means of the following multiplicative
identity  in the generating functional \cite{HubbStra}
\begin{equation}
\label{Hubident}
1=N'\int D\Psi\exp\{i\int d^{4}x\frac{4!}{\lambda}[\Psi+\frac{\lambda}{4!}(\tilde{\phi}^{2}+2\phi_{0}\tilde{\phi})]^{2}\}
\end{equation}
{wherein} $N'$ is a normalization, $\Psi$ are the auxiliary  field and it has a wave function renormalization factor ${\cal Z}_{\Psi}$.
It  yields the following form for the generating functional:
\begin{eqnarray}
\label{eqpsi}
&&Z=N\int D\bar{q}DqD\tilde{\phi}D\Psi\exp[i\int d^{4}x\{\bar{q}i\slashed{\partial}q-g(\phi_{0}+\tilde{\phi})\bar{q}q+\cr\cr
&&-\Box\phi_{0}\tilde{\phi}+2m^{2}\phi_{0}\tilde{\phi}-\frac{\lambda}{3!}\phi_{0}^{3}\tilde{\phi}+4\phi_{0}\Psi\tilde{\phi}+\cr\cr
&&-\frac{1}{2}\tilde{\phi}(\Box-2 m^{2}-4\Psi-4\frac{\lambda}{4!}\phi_{0}^{2})\tilde{\phi}+\frac{4!}{\lambda}\Psi^{2} + {\cal L}_s 
\}\exp(i\Gamma_{0}),
\end{eqnarray}
in which $\Gamma_{0}$ is the effective action that collected all the terms
exclusively  associated with the background
field and that is given by:
\begin{equation} \label{gamma0}
\Gamma_{0}=\int d^{4}x[\frac{1}{2}(\partial_{\mu}\phi_{0})^{2}
+ {  m^{2} }
\phi_{0}^{2}-\frac{\lambda}{4!}\phi_{0}^{4}].
\end{equation}

To make possible a latter 
current expansion for
 eq. (\ref{eqpsi}), the scalar field can be 
exactly integrated out by means of  the following field translation:
\begin{eqnarray}
&&\tilde{\phi}(x)\rightarrow\tilde{\phi}_{0}(x)+\int d^{4}yG(x,y)j(y)\cr\cr
\mbox{where}&&j=-g\bar{q}q-\Box\phi_{0}+2m^{2}\phi_{0}-\frac{\lambda}{3!}\phi_{0}^{3}+4\phi_{0}\Psi,
\cr\cr
\mbox{and} && 
G^{-1}(x,y)=(\Box-2 m^{2}-4\Psi-4\frac{\lambda}{4!}\phi_{0}^{2})\delta^{4}(x-y).
\end{eqnarray}
It is interesting to note that: $ -\frac{1}{2}\tilde{\phi}G^{-1}\tilde{\phi}+j\tilde{\phi}=
-\frac{1}{2}\tilde{\phi}_{0}G^{-1}\tilde{\phi}_{0}+\frac{1}{2}\; j \; G \; j$.

The functional generator can then be written as:
\begin{eqnarray}  \label{Z-for-phi0-int}
Z &=& 
N\int D\bar{q}DqD\tilde{\phi}_{0}D\Psi\exp[i\int d^{4}x\{\bar{q}i\slashed{\partial}q-g\phi_{0}\bar{q}q
\cr\cr
 &+&  \frac{4!}{\lambda}\Psi^{2}
+ \int d^4 y \left( - \frac{1}{2}\tilde{\phi}_{0}G^{-1}\tilde{\phi}_{0}+\frac{1}{2}jGj \right)
+ {\cal L}_s \}\exp(i\Gamma_{0}).
\end{eqnarray}
Therefore it is possible to define a mass for the scalar field in both cases, when
the fields $\phi_0$ and $\Psi$ develop or not {a} non zero expected value 
in the vacuum  as discussed below.
In general the scalar field mass for $\tilde{\phi}_{0}$, by assuming the possibility of  non trivial classical 
solutions,   can be written, except for renormalization effects shown below in 
Section (\ref{sec:renormaliza-gaps}), as:
\begin{equation}
\label{Mphi}
{M}^2_\phi =2 m^2 + 4({\Psi}_0+\frac{\lambda}{4!} \phi_0^2 ).
\end{equation}
In this equation we see the contribution of the SSB to the scalar particle mass by means of $\phi_0$ 
and the contribution from an eventual contribution of the expected value in the vacuum
of the auxiliary field $\Psi_0$ in the case its gap equation, discussed below, presents
non trivial solution. It corresponds to consider higher order contribution for the usual mean field
solution of the SSB for $\phi$.
The composite field $\Psi$ will be analyzed below.
In this equation there is a contribution from the spontaneous symmetry breaking ($\phi_0$) 
and another from the  auxiliary field,
whenever it develops a non zero classical value in the vacuum.
 The auxiliary  scalar field $\Psi$ corresponds to a two boson quantum state and  its dynamics is completely undetermined so far. 
With the identity $\det A=\exp[tr\; \ln\; A]$ 
 we write the effective action as it follows:
\begin{eqnarray}
\label{firstpsi0}  
&&   S_{eff}= - \frac{i}{2}Tr \; \ln[G^{-1}(x,y)] + \int d^4 x \; \frac{4!}{\lambda}\Psi(x)^{2},
\end{eqnarray}
and the corresponding extremization that
 lead to a gap equation,   hopefully defining a  ground state.
It will be given by:
\begin{eqnarray} 
\label{gap-psi} 
&& 
\frac{\partial S_{eff}}{\partial \Psi}|_{\Psi=\Psi_{0}}=-
\frac{i}{
[\Box-M_{\phi}^{2}]}\delta^{4}(x-y)
+\frac{4!}{\lambda}\Psi_{0}=0,
\end{eqnarray}
and so, in the momentum representation we have that
\begin{eqnarray} 
\label{gap-psi-m}
\Psi_{0}=\frac{\lambda}{4!}\int\frac{d^{4}p}{(2\pi)^{4}}\frac{i}{
[p^{2}+2 m^2 + 4({\Psi}_{0}+\frac{\lambda}{4!} \phi_0^2 )]}.
\end{eqnarray}
This might yield a non trivial solution  $\Psi_{0}$ that eventually  contributes  for the 
effective mass of the scalar field $\tilde{\phi}_{0}$ of {the} expression (\ref{Mphi}).
 This equation has a quadratic divergence, typical from gap equations.

\subsubsection{ Higgs-type SSB}

Above it was assumed that the scalar field might develop a classical value in the vacuum, as discussed for the tree level, {in} eq. (\ref{ssb-tree}). Therefore  it is {also} relevant to consider again the equation {that define} the expected value of the scalar field $\phi_0$ with quantum corrections.
By extremizing eq. (\ref{gamma0}) with the 
one loop corrections (\ref{firstpsi0}) with respect to it. 
Then a corrected gap equation arises and it is given by:
\begin{eqnarray}
&&\frac{\partial S_{eff}}{\partial \phi}|_{\phi_{0}} \to
\left(m^{2}-\frac{\lambda}{{12} }\phi_{0}^{2}
+\frac{i\lambda}{6[\Box-M_{\phi}^{2}]}
\delta^{4}(x-y) \right) 
{2}\phi_{0} = 0.
\end{eqnarray}
The solutions for this equation can be written as:
\begin{eqnarray}
\label{SSBsolution}
\phi_0 = 0, \;\;\;\;\;
\phi_0^{2} = 
\left(\frac{{12} m^2 }{\lambda}
+ \frac{{2}i}{[\Box - M_{\phi}^2]} \delta^4 (x-y) \right),
\end{eqnarray}
{in which the non-trivial solutions can be written in the momentum representation
\begin{eqnarray}
\label{SSBsolution-m}
\phi_0^{2} =\frac{{12} m^2 }{\lambda}
-
\int\frac{d^{4}p}{(2\pi)^{4}}\frac{2i}{p^{2}+ 2 m^2 + 4({\Psi}_{0}+\frac{\lambda}{4!}\phi_0^2)}.
\end{eqnarray}}
By considering eq. (\ref{gap-psi-m})
this last eq. can be written as:
\begin{eqnarray}  \label{ssb-phipsi}
\phi_0^{2}=\bar{\phi}_0^{2}
- \frac{48}{\lambda} \Psi_0.
\end{eqnarray}
This equation makes explicit the role of the auxiliary field $\Psi$ for the
 SSB.
 Note that, when computing the effective action for $\phi_0$  in Section (\ref{Sec12})
a spacetime dependence will be taken into account.

\subsubsection{Renormalized SSB gap equations}
\label{sec:renormaliza-gaps}

Now we renormalize the previous gap equations
in eq. (\ref{gap-psi-m}) and eq. (\ref{SSBsolution}) respectively.
 By means of eq. (\ref{Lr}) and applying the following renormalized (HS) transformation
\begin{equation}
1=N'\int D\Psi_{R}\exp\{i\int d^{4}x\frac{4!}{\lambda_{R}}[{\cal Z}_{\Psi}^{\frac{1}{2}}\Psi_{R}+{\cal Z}_{\lambda}^{\frac{1}{2}}\frac{\lambda_{R}}{4!}(\tilde{\phi}^{2}_{R}+2\phi_{0}\tilde{\phi}_{R})]^{2}\}.
\end{equation}
we have the general result
\begin{eqnarray}
\label{gapr1}
&&\phi_0^{2}(R,c.t)=\frac{{\cal Z}_m}{{\cal Z}_\lambda}\frac{12m_R^2}{\lambda_R}
-\frac{{\cal Z}_{\Psi}^{\frac{1}{2}}}{{\cal Z}_{\lambda}^{\frac{1}{2}}}\frac{48\Psi_{0}(R,c.t)}{\lambda_{R}}\cr\cr
&&\Psi_{0}(R,c.t)=\frac{i{\cal Z}_{\lambda}^{\frac{1}{2}}\lambda_{R}}{24{\cal Z}_{\Psi}^{\frac{1}{2}}[{\cal Z}_{\phi}\Box-M_{\phi}^{2}(R,c.t)]}\delta^{4}(x-y)\cr\cr
&&{M}^2_\phi(R,c.t)=2{\cal Z}_m m^2_R + 4[{\cal Z}_\Psi^\frac{1}{2}
 {\cal Z}_{\lambda}^{\frac{1}{2}}
{\Psi}_{0}(R,c.t) + {\cal Z}_\lambda \frac{\lambda_R}{4 !}\phi_0^2(R,c.t)],
\end{eqnarray}
wherein the notation (R,c.t) is to remind us that now the objects are function of dressed quantities (R) and counter-terms (c.t) due to the renormalization factors ${\cal Z}_{i}$. The first equation, for ${\phi}_0(R,c.t)$, presents  a quadratic divergence in the correction 
 that is  renormalized by ${\cal Z}_m$ and compatible with a finite mass condition for $m_{R}$.
 Although that equation looks like the classical $\lambda \phi^4$ gap equation
it is a highly non linear equation  -
and this is seen in the second of these equations - which reduces to the 
classical level equation by setting the one loop contribution to zero.
The effective mass $M_\phi^2$ dependence on $\phi_0$, however, 
 introduces further non linearities not only 
because of its dependence on $\phi_0$ but also because it depends
on the auxiliary field expected value $\Psi_0$.
At this point it is important to note that, the pole of the two point function of $\tilde{\phi}_{0}$ is 
real and positive, $M_\phi^2(R,c.t)>0$.

\subsection{Current expansion and fermion-effective action}

Consider the non linear term in eq. (\ref{Z-for-phi0-int}), which depends on the fermion 
current $j$,
and that can be written as:
\begin{eqnarray}
&& \frac{1}{2}\int d^{4}x d^{4}yjGj =
-\frac{1}{2}\int d^{4}x[g\bar{q}(x)q(x)-\Box\phi_{0}+2m^{2}\phi_{0}-\frac{\lambda}{3!}\phi_{0}^{3}+4\phi_{0}\Psi(x)]\times \cr\cr
&& \int d^{4}y\int\frac{d^{4}p}{(2\pi)^{4}}\frac{[g\bar{q}(y)q(y)-\Box\phi_{0}+2m^{2}\phi_{0}-\frac{\lambda}{3!}\phi_{0}^{3}+4\phi_{0}\Psi(y)]}{[p^{2}+M_{\phi}^2]}\exp[ip(x-y)],
\end{eqnarray}
We now assume that the kinetic part of the scalar field is suppressed by the total
(large)  mass term {($p^{2}\ll M_{\phi}^{2}$), in a way similar to the relation between the electroweak theory and the effective Fermi theory due to the large mass of the interaction mediators}.
 In this case, the following local limit can be taken:
\begin{eqnarray}
\frac{1}{2}\int d^{4}x d^{4}yjGj &\cong &-\frac{1}{2}\int d^{4}x
[a(\bar{q}q)^{2}+ {2} b(\bar{q}q)+
 c\Psi+d\Psi^{2}+f]
\end{eqnarray}
wherein
\begin{eqnarray}
&&a=\frac{g^{2}}{M_{\phi}^{2}},
\cr\cr
&&
 b=\frac{g}{M_{\phi}^{2}}[2m^{2}\phi_{0}-\frac{\lambda}{3!}\phi_{0}^{3}+4\phi_{0}\Psi],
\cr\cr
&&
c=\frac{8\phi_{0}}{M_{\phi}^{2}}[2m^{2}\phi_{0}-\frac{\lambda}{3!}\phi_{0}^{3}],
\cr\cr
&&
 d=\frac{16\phi_{0}^{2}}{M_{\phi}^{2}}
,
\cr\cr
&&f=\frac{[2m^{2}\phi_{0}-\frac{\lambda}{3!}\phi_{0}^{3}]^{2}}{M_{\phi}^{2}},
\end{eqnarray}
The following expression is obtained for the effective action of the model:
\begin{eqnarray}
\label{eqS}
&&Z=N\int D\bar{q}DqD\tilde{\phi}_{0}D\Psi\exp[i\int d^{4}x\{\bar{q}i\slashed{\partial}q-g\phi_{0}\bar{q}q-{b} \bar{q}q+C_{F}(\bar{q}q)^{2}+\cr\cr
&&-\frac{1}{2}\tilde{\phi}_{0}G^{-1}\tilde{\phi}_{0}- \frac{c}{2}
\Psi+[\frac{4!}{\lambda}-\frac{d}{2}]\Psi^{2}-\frac{f}{2}\}\exp(i\Gamma_{0}),
\end{eqnarray}
in which we could define the renormalized Fermi constant 
$C_{F}^{R}=-\frac{1}{2}\frac{g_{R}^{2}}{M_{\phi}^{2}}$ with ${\cal Z}_{C_{F}}={\cal Z}_{g}^{2}$. \footnote{In the regime where $\Psi_{0}$ suppresses the other masses we have that $C_{F}^{R}=-\frac{1}{4}\frac{g_{R}^{2}}{\Psi_{0}}$.}
 Also note that,
 as $\phi_{0}$ is a  constant, { or  very slowly varying}, background,
we can set $\Box\phi_{0}=0$.

\subsection{Coupled one loop gap equations}

Fermion degrees of freedom were kept so far intact
and now the AFM is considered again to reduce
the fermion self interactions
 into bilinears. 
Before doing  that however, firstly
let us  introduce  a background fermionic current
by  shift of the bilinear, that is needed for the one loop calculation,
  $(\bar{q}q)_{0}$ as
 $\bar{q}q\rightarrow(\bar{q}q)_{0}+(\bar{q}q) $ 
in which $(\bar{q}q) $ is the quantum field.
The auxiliary field  for the fluctuations are introduced again
by means of the following unit integral in the generating functional:
\begin{equation}
1=N''\int DS\exp[-i\int d^{4}x\frac{1}{4C_{F}}(S+2C_{F}({\bar{q}}{q}))^{2}].
\end{equation}
We  are left with the following functional generator:
\begin{eqnarray}
\label{bigeq}
&&Z= N\int D \bar{q} D {q}D\tilde{\phi}_{0}DS D\Psi
\exp(i\Gamma_{0})
\exp[i\int d^{4}x\{\bar{q}(i\slashed{\partial}-g\phi_{0}-{b }-S+ 2C_{F}(\bar{q}q)_{0})q+
\cr\cr
&&-\frac{1}{4C_{F}}S^{2}-\frac{1}{2}\tilde{\phi}_{0}(\Box-2 m^{2}-4\Psi-4\frac{\lambda}{4!}\phi_{0}^{2})\tilde{\phi}_{0}- \frac{c}{2}
\Psi+[\frac{4!}{\lambda}-\frac{d}{2}]\Psi^{2}+\frac{1}{4C_{F}}S^{2}-\frac{f}{2}
+ \Gamma_{q_0}\},
\end{eqnarray}
{ where 
$\Gamma_{q_0}$ contains the  other  terms that depend exclusively on the background fermion field 
$(\bar{q} q)_0$.}
It is possible to define a total dressed mass for the fermion field in the case that the fields $\phi_0$ and $S$ develop classical solutions in  the corresponding gap equations.
This effective mass can be written as
\begin{equation}
\label{MS}
M_{q}=g\phi_{0}+ {b} +S,
\end{equation}
wherein each of the terms has a precise physical meaning:
 the first represents a SSB of the Higgs-type, 
{{
the second ($b$) contains different contributions from the Higgs-type order parameter, $\phi_0$, and 
a correction to the Higgs mechanism due to two-scalars correlations in the vacuum, $\Psi$,
and, at 
the last,  a dynamical symmetry breaking by means of $S$.
}

By  proceeding  with the integration in the fluctuation fields we have:
\begin{eqnarray}
\label{finaleq}
&&Z=N\int DSD\Psi\det[S_F^{-1}]\det[S_B^{-1}]^{-\frac{1}{2}}\cr\cr
&&\exp[i\int d^{4}x\{\frac{1}{4C_{F}}S^{2}- \frac{c}{2}
\Psi+[\frac{4!}{\lambda}-\frac{d}{2}]\Psi^{2}-\frac{f}{2}\}\exp(i\Gamma_{0})
\end{eqnarray}
where
{
\begin{eqnarray}
&&S_F(x,y) = (i\slashed{\partial}-M_{q})^{-1} \delta^4(x-y) ,
\cr\cr
&&S_B(x,y) = (\Box-M_{\phi}^{2} )^{-1} \delta^4(x-y).
\end{eqnarray}}

To calculate the full set of (gap) equations that define the ground state of the system,
let us write 
 the effective potential $V_{eff}$ from eq. (\ref{finaleq}).
{By means of the identity $\det A=\exp[tr\ln A]$ we write
\begin{eqnarray}
-V_{eff}&=& -itr\int d^{4}y\ln[S_{F}(x,y)]\delta^{4}(x-y)+\frac{1}{4C_{F}}S^{2}(x)+\frac{i}{2}\int d^{4}y\ln[S_{B}(x,y)]\delta^{4}(x-y)+\cr\cr
&&-
\frac{c}{2} \Psi(x)+[\frac{4!}{\lambda}-\frac{d}{2}]\Psi^{2}(x)-\frac{f}{2}.
\end{eqnarray}}
By extremizing this effective potential with respect to the auxiliary fields $\Psi$ and $S$
the following gap equations are obtained:
\begin{eqnarray}
\label{gapfinaleq}
\frac{\partial V_{eff}}{\partial \Psi}|_{\Psi=\Psi_{0}}&=&
itr\frac{2}{[i\slashed{\partial}-M_{q}]}\frac{[b+g\phi_{0}]}{M^{2}_{\phi}}\delta^{4}(x-y)
-\frac{2}{g^{2}}S_{0}^{2}\cr\cr
&&-i\frac{2}{[\Box-M_{\phi}^{2}]}\delta^{4}(x-y)+(\frac{2\Psi_{0}}{M_{\phi}^{2}}- \frac{1}{2}
)c+2[\frac{4!}{\lambda}+(\frac{4\Psi_{0}}{M_{\phi}^{2}}-\frac{1}{2})d]\Psi_{0}
+\frac{2f}{M^{2}_{\phi}}=0,
\cr\cr
\label{gapfinaleq2}
\frac{\partial V_{eff}}{\partial S}|_{S=S_{0}}&=&
itr\frac{1}{[i\slashed{\partial}-M_{q}]}\delta^{4}(x-y)+\frac{1}{2C_{F}}S_{0}=0,
\end{eqnarray}
{ where the first of these equations
 provides corrections to the eq. (\ref{gap-psi})
and 
the second equation provides the usual DChSB.
A non trivial solution  for the first equation ($\Psi_0$) leads to 
a contribution for the scalar boson mass and a redefinition of the scalar field 
condensate (SSB) (\ref{ssb-phipsi}).
}
 It can be noted that this pair of equations, together with the
quantum gap eq. for $\phi_0$,
provides further account of YuM interactions than usual one loop equation derived for a single
auxiliary field.

\subsubsection{Renormalized coupled one loop gap equations}

{Now let us  renormalize the coupled gap equations in eq. (\ref{gapfinaleq}) and eq. (\ref{gapfinaleq2}), by means of eq. (\ref{Lr}) by  applying the following renormalized (HS) transformation:
\begin{eqnarray}
&&1=N'\int D\Psi_{R}\exp\{i\int d^{4}x\frac{4!}{\lambda_{R}}[{\cal Z}_{\Psi}^{\frac{1}{2}}\Psi_{R}+{\cal Z}_{\lambda}^{\frac{1}{2}}\frac{\lambda_{R}}{4!}(\tilde{\phi}^{2}_{R}+2\phi_{0}\tilde{\phi}_{R})]^{2}\}\cr\cr
&&1=N''\int DS_{R}\exp[-i\int d^{4}x\frac{1}{4C_{F}^{R}}({\cal Z}_{S}^{\frac{1}{2}}S_{R}+2{\cal Z}_{C_{F}}^{\frac{1}{2}}C_{F}^{R}( {\bar{q}}_{R} {q}_{R}))^{2}].
\end{eqnarray}
We  are left with the following functional generator
\begin{eqnarray}
\label{bigeqR}
Z &= & N\int D \bar{q}_{R} D {q}_{R}D\tilde{\phi}_{0}DS_{R}D\Psi_{R}
\exp(i\Gamma_{0})
\exp[i\int d^{4}x
\{
\cr\cr
&&
 {\bar{q}}_{R}({\cal Z}_{q}i\slashed{\partial}-{\cal Z}_{g}g_{R}\phi_{0}-\frac{b(R,c.t)}{2}-{\cal Z}_{S}^{\frac{1}{2}}{\cal Z}_{C_{F}}^{\frac{1}{2}}S_{R}+ 2{\cal Z}_{C_{F}}C_{F}^{R}(\bar{q}q)_{0})
 {q}_{R}
\cr\cr
&&
-\frac{1}{4C_{F}^{R}}{\cal Z}_{S}S_{R}^{2}-\frac{1}{2}\tilde{\phi}_{0}({\cal Z}_{\phi}\Box-2{\cal Z}_{m}m_{R}^{2}-4{\cal Z}_{\Psi}^{\frac{1}{2}}{\cal Z}_{\lambda}^{\frac{1}{2}}\Psi_{R}-4{\cal Z}_{\lambda}\frac{\lambda_{R}}{4!}\phi_{0}^{2})\tilde{\phi}_{0}\cr\cr
&&-\frac{c(R,c.t)}{2}\Psi_{R}+[\frac{4!}{\lambda_{R}}{\cal Z}_{\Psi}-\frac{d(R,c.t)}{2}]\Psi_{R}^{2}+\frac{1}{4C_{F}^{R}}{\cal Z}_{C_{F}}S_{R}^{2}-\frac{f(R,c.t)}{2}\},
\end{eqnarray}
and from it we find the following set of renormalized gap equations:

\begin{eqnarray}
\label{gapr2}
(\Psi_0:) &&itr\frac{2{\cal Z}_{\Psi}^{\frac{1}{2}}{\cal Z}_{\lambda}^{\frac{1}{2}}}{[{\cal Z}_{q}i\slashed{\partial}-M_{q}(R,c.t)]}\frac{[b(R,c.t)+{\cal Z}_{g}g_{R}\phi_{0}]}{M^{2}_{\phi}(R,c.t)}\delta^{4}(x-y)-\frac{2{\cal Z}_{\Psi}^{\frac{1}{2}}{\cal Z}_{\lambda}^{\frac{1}{2}}{\cal Z}_{C_{F}}}{g_{R}^{2}}S_{0}^{2}+\cr\cr
&&-i\frac{2{\cal Z}_{\Psi}^{\frac{1}{2}}{\cal Z}_{\lambda}^{\frac{1}{2}}}{[{\cal Z}_{\phi}\Box-M_{\phi}^{2}(R,c.t)]}\delta^{4}(x-y)
+ (\frac{2{\cal Z}_{\Psi}^{\frac{1}{2}}{\cal Z}_{\lambda}^{\frac{1}{2}}\Psi_{0}}{M_{\phi}^{2}(R,c.t)}- \frac{1}{2}
)c(R,c.t)+\cr\cr
&&+2[\frac{4!}{\lambda_{R}}{\cal Z}_{\Psi}+(\frac{4{\cal Z}_{\Psi}^{\frac{1}{2}}{\cal Z}_{\lambda}^{\frac{1}{2}}\Psi_{0}}{M_{\phi}^{2}(R,c.t)}-\frac{1}{2})d(R,c.t)]\Psi_{0}+\frac{2{\cal Z}_{\Psi}^{\frac{1}{2}}{\cal Z}_{\lambda}^{\frac{1}{2}}f(R,c.t)}{M^{2}_{\phi}(R,c.t)}=0,\cr\cr
(S_0:)&&itr\frac{{\cal Z}_{S}^{\frac{1}{2}}{\cal Z}_{C_{F}}^{\frac{1}{2}}}{[{\cal Z}_{q}i\slashed{\partial}-M_{q}(R,c.t)]}\delta^{4}(x-y)+\frac{1}{2C_{F}^{R}}{\cal Z}_{C_{F}}S_{0}=0,
\end{eqnarray}
in which
\begin{eqnarray} \label{Mphiren}
&&{M}^2_\phi(R,c.t)=2{\cal Z}_m m^2_R + 4({\cal Z}_\Psi^\frac{1}{2}
 {\cal Z}_{\lambda}^{\frac{1}{2}}
{\Psi}_{R} + {\cal Z}_\lambda \frac{\lambda_R}{4 !}\phi_0^2)\cr\cr
\label{Mqren}
&&M_{q}(R,c.t)={\cal Z}_{g}g_{R}\phi_{0}+
b(R,c.t)+{\cal Z}_{S}^{\frac{1}{2}}{\cal Z}_{C_{F}}^{\frac{1}{2}}S_{R},
\end{eqnarray}
and 
\begin{eqnarray} 
&&
b(R,c.t)=\frac{{\cal Z}_{g}g_{R}}{M_{\phi}^{2}(R,c.t)}[2Z_{m}m_{R}^{2}\phi_{0}-{\cal Z}_{\lambda}\frac{\lambda_{R}}{3!}\phi_{0}^{3}+4{\cal Z}_{\Psi}^{\frac{1}{2}}{\cal Z}_{\lambda}^{\frac{1}{2}}\phi_{0}\Psi_{R}],
\cr\cr
&&
c(R,c.t)=\frac{8{\cal Z}_{\Psi}^{\frac{1}{2}}{\cal Z}_{\lambda}^{\frac{1}{2}}\phi_{0}}{M_{\phi}^{2}(R,c.t)}[2{\cal Z}_{m}m_{R}^{2}\phi_{0}-{\cal Z}_{\lambda}\frac{\lambda_{R}}{3!}\phi_{0}^{3}],
\cr\cr
&&
d(R,c.t)=\frac{16{\cal Z}_{\Psi}{\cal Z}_{\lambda}\phi_{0}^{2}}{M_{\phi}^{2}(R,c.t)}
,
\cr\cr
&&f(R,c.t)=\frac{[2{\cal Z}_{m}m_{R}^{2}\phi_{0}-{\cal Z}_{\lambda}\frac{\lambda_{R}}{3!}\phi_{0}^{3}]^{2}}{M_{\phi}^{2}(R,c.t)}.
\end{eqnarray}}

There are  two mass generation mechanisms in play, dynamic mass generation due to the 
 dynamical chiral symmetry  breaking
and the mass generated by the spontaneous symmetry breaking due to the $Z_{2}$ discrete symmetry. As we see in eq. (\ref{Lagrern}) we can 
associate two physical conditions for the fermion sector and three physical conditions for the scalar sector (propagators and vertices). So, with five counter-terms we adjust the theory.
Actually we have three coupled gap equations, $\phi_0$ and $\Psi_0$ for the boson and 
$S_0$ for the fermions. All of them can contribute for both boson and fermion masses 
according to eqs. (\ref{Mphiren}).
 On the other hand, when the dynamic mass generation happens, we have one gap associated with the fermion mass and other 
 associated with the scalar mass, and the gap coupled equations are divergent. Therefore, in addition to the regularization process \cite{Polchi} (cut-off, for example), apparently we need two more counter-terms to extract a physical gap.

It is not clear whether these coupled gap equations  might have an unique solution or not. To make sure these are solutions of minimum action a second derivative might be taken. Therefore, we write the Hessian matrix from the renormalized effective action in eq. (\ref{bigeqR})
\begin{equation}
\label{H}
[H]=\begin{pmatrix}
\frac{\partial^{2}S_{eff}^{R}}{\partial\Psi\partial\Psi}&\frac{\partial^{2}S_{eff}^{R}}{\partial\Psi\partial S}\\
\frac{\partial^{2}S_{eff}^{R}}{\partial S\partial\Psi}&\frac{\partial^{2}S_{eff}^{R}}{\partial S\partial S}
\end{pmatrix}
\end{equation}
in which we have the following terms

\begin{eqnarray}
&&\frac{\partial^{2}S_{eff}^{R}}{\partial\Psi\partial\Psi}=u_{1}+u_{2}+u_{3}+u_{4}+u_{5}\cr\cr
&&u_{1}=itr\frac{8{\cal Z}_{\Psi}{\cal Z}_{\lambda}}{[{\cal Z}_{q}i\slashed{\partial}-M_{q}(R,c.t)]^{2}}\frac{[b(R,c.t)+{\cal Z}_{g}g_{R}\phi_{0}]^{2}}{M^{4}_{\phi}(R,c.t)}\delta^{4}(x-y)-itr\frac{8{\cal Z}_{\Psi}{\cal Z}_{\lambda}}{[{\cal Z}_{q}i\slashed{\partial}-M_{q}(R,c.t)]}\frac{b(R,c.t)}{M^{4}_{\phi}(R,c.t)}\delta^{4}(x-y)+\cr\cr
&&-itr\frac{8{\cal Z}_{\Psi}{\cal Z}_{\lambda}}{[{\cal Z}_{q}i\slashed{\partial}-M_{q}(R,c.t)]}\frac{b(R,c.t)+{\cal Z}_{g}g_{R}\phi_{0}}{M^{4}_{\phi}(R,c.t)}\delta^{4}(x-y)\cr\cr
&&u_{2}=itr\frac{8{\cal Z}_{\Psi}{\cal Z}_{\lambda}}{[{\cal Z}_{q}i\slashed{\partial}-M_{q}(R,c.t)]^{2}}\delta^{4}(x-y)\cr\cr
&&u_{3}=(\frac{2{\cal Z}_{\Psi}^{\frac{1}{2}}{\cal Z}_{\lambda}^{\frac{1}{2}}}{M_{\phi}^{2}(R,c.t)}-\frac{8{\cal Z}_{\Psi}^{\frac{1}{2}}{\cal Z}_{\lambda}^{\frac{1}{2}}\Psi_{0}}{M^{4}_{\phi}}
)c(R,c.t)-(\frac{2{\cal Z}_{\Psi}^{\frac{1}{2}}{\cal Z}_{\lambda}^{\frac{1}{2}}\Psi_{0}}{M_{\phi}^{2}(R,c.t)}- \frac{1}{2}
)\frac{4{\cal Z}_{\Psi}^{\frac{1}{2}}{\cal Z}_{\lambda}^{\frac{1}{2}}c(R,c.t)}{M^{2}_{\phi}}\cr\cr
&&u_{4}=2(\frac{4{\cal Z}_{\Psi}^{\frac{1}{2}}{\cal Z}_{\lambda}^{\frac{1}{2}}}{M_{\phi}^{2}(R,c.t)}-\frac{16{\cal Z}_{\Psi}{\cal Z}_{\lambda}\Psi_{0}}{M^{4}_{\phi}})d(R,c.t)\Psi_{0}-(\frac{4{\cal Z}_{\Psi}^{\frac{1}{2}}{\cal Z}_{\lambda}^{\frac{1}{2}}\Psi_{0}}{M_{\phi}^{2}(R,c.t)}-\frac{1}{2})\frac{32{\cal Z}_{\Psi}{\cal Z}_{\lambda}d(R,c.t)}{M^{2}_{\phi}}\Psi_{0}+\cr\cr
&&+2[\frac{4!}{\lambda_{R}}{\cal Z}_{\Psi}+(\frac{4{\cal Z}_{\Psi}^{\frac{1}{2}}{\cal Z}_{\lambda}^{\frac{1}{2}}\Psi_{0}}{M_{\phi}^{2}(R,c.t)}-\frac{1}{2})d(R,c.t)]\cr\cr
&&u_{5}=-\frac{16{\cal Z}_{\Psi}^{\frac{1}{2}}{\cal Z}_{\lambda}^{\frac{1}{2}}f(R,c.t)}{M^{2}_{\phi}(R,c.t)},
\end{eqnarray}

\begin{equation}
\frac{\partial^{2}S_{eff}^{R}}{\partial\Psi\partial S}=itr\frac{{\cal Z}_{\Psi}^{\frac{1}{2}}{\cal Z}_{\lambda}^{\frac{1}{2}}{\cal Z}_{S}^{\frac{1}{2}}{\cal Z}_{C_{F}}^{\frac{1}{2}}}{[{\cal Z}_{q}i\slashed{\partial}-M_{q}(R,c.t)]^{2}}\frac{[b(R,c.t)+{\cal Z}_{g}g_{R}\phi_{0}]}{M^{2}_{\phi}(R,c.t)}\delta^{4}(x-y)-\frac{2{\cal Z}_{\Psi}^{\frac{1}{2}}{\cal Z}_{\lambda}^{\frac{1}{2}}{\cal Z}_{C_{F}}S_{0}}{g^{2}_{R}},
\end{equation}

\begin{equation}
\frac{\partial^{2}S_{eff}^{R}}{\partial S\partial\Psi}=itr\frac{{\cal Z}_{\Psi}^{\frac{1}{2}}{\cal Z}_{\lambda}^{\frac{1}{2}}{\cal Z}_{S}^{\frac{1}{2}}{\cal Z}_{C_{F}}^{\frac{1}{2}}}{[{\cal Z}_{q}i\slashed{\partial}-M_{q}(R,c.t)]^{2}}\frac{[b(R,c.t)+{\cal Z}_{g}g_{R}\phi_{0}]^{2}}{M^{4}_{\phi}(R,c.t)}\delta^{4}(x-y),
\end{equation}

\begin{equation}
\frac{\partial^{2}S_{eff}^{R}}{\partial S\partial S}=itr\frac{{\cal Z}_{S}{\cal Z}_{C_{F}}}{[{\cal Z}_{q}i\slashed{\partial}-M_{q}(R,c.t)]^{2}}\delta^{4}(x-y)+\frac{1}{2C_{F}^{R}}{\cal Z}_{C_{F}}.
\end{equation}
So the condition of local minimum can be achieved for the extremal points ($\Psi_{0},S_{0}$) if the characteristic equation $det([H]-\nu I)$ has positive eigenvalues ($\nu_{\pm}$).
In the large mass limit for the intermediate boson ($M_{\phi}$)
 the off-diagonal terms of the Hessian do not contribute and we have following conditions
\begin{eqnarray}
&&itr\frac{8{\cal Z}_{\Psi}{\cal Z}_{\lambda}}{[{\cal Z}_{q}i\slashed{\partial}-M_{q}(R,c.t)]^{2}}\delta^{4}(x-y)+2(\frac{4!}{\lambda_{R}}{\cal Z}_{\Psi}-\frac{1}{2})d(R,c.t)]>0\cr\cr
&&itr\frac{{\cal Z}_{S}{\cal Z}_{C_{F}}}{[{\cal Z}_{q}i\slashed{\partial}-M_{q}(R,c.t)]^{2}}\delta^{4}(x-y)+\frac{1}{2C_{F}^{R}}{\cal Z}_{C_{F}}>0.
\end{eqnarray}

\subsubsection{Coupled gap  equations for  particular limits 
of $m$, $\lambda$ and $g$}

As we discussed previously, 
to solve the previous  coupled
gap  equations in eq. (\ref{gapr1}) and eq. (\ref{gapr2}) may  not be 
an easy task, due to the arbitrariness of couplings and masses along with the concept of renormalization.
 Below some particular prescriptions for solving the gap equations are 
presented by considering the usual logics of low energy effective models since
the resulting model in terms of auxiliary fields is non renormalizable.
Formally, we have 3 equations and 3 gaps ($\phi_0, \Psi_{0}, S_{0}$), that can be written momentum representation as follows (${\cal Z}_{i} \rightarrow 1$):

\begin{eqnarray}
\label{gaprc}
\phi_0^{2} &=& \frac{12 m^2 }{\lambda}
-\int\frac{d^{4}p}{(2\pi)^{4}}\frac{2i}{p^{2}+M_{\phi}^{2}},\cr\cr
[\frac{2\Psi_{0}}{M_{\phi}^{2}}- \frac{1}{2}]c
 &=&-8i\frac{[b+g\phi_{0}]}{M_{\phi}^{2}}\int\frac{d^{4}p}{(2\pi)^{4}}\frac{M_{q}}{p^{2}+M_{q}^{2}}-\frac{2}{g^{2}}S_{0}^{2}+2i\int\frac{d^{4}p}{(2\pi)^{4}}\frac{i}{p^{2}+M_{\phi}^{2}}
\cr\cr
&&+2[\frac{4!}{\lambda}+(\frac{4\Psi_{0}}{M_{\phi}^{2}}-\frac{1}{2})d]\Psi_{0}+\frac{2f}{M_{\phi}^{2}},
\cr\cr
S_{0}&=& 
-8C_{F}i\int\frac{d^{4}p}{(2\pi)^{4}}\frac{\bar{M}_{q}}{[ p^{2}- \bar{M}_{q}^{2}]}.
\end{eqnarray}

Thus, we will explore solutions of the gap coupled equations in the following limits:
\begin{eqnarray*}
&&\text{a)$\lambda\phi_{0}^{2}\sim m^{2}$, $\phi_{0}\neq0$\quad (solution for SSB without quantum contributions)}\cr\cr
&&\text{b)$m\rightarrow\infty$ ,\quad (scalar particle, mediator of the interaction, with large mass)}\cr\cr
&&\text{c)$g \phi_0 \sim m$}\quad (\text{"ad doc" constrain in the Yukawa coupling}).
\end{eqnarray*}
So that, the massive particle
 associated with the $\phi$ is in some sense localized.
We have as consequences the results below
\begin{eqnarray}
&&g^{2}\sim \lambda,\cr\cr
&&C_{F}\sim-\frac{g^{2}}{m^{2}}\sim -\frac{1}{\phi_{0}^{2}}.
\end{eqnarray}
Therefore the dynamical mass mechanism are dictated by the value of $\phi_{0}$ and the SSB. 
As a consequence,
 in the weak regime for the couplings ($g$, $\lambda$), $C_{F}\rightarrow0$ and 
$\phi_{0}^{2}\rightarrow\infty$. In another hand, in the 
regime where the couplings ($g$, $\lambda$) are not too weak, or in another words,  stronger, $C_{F}<0$ and $\phi_{0}^{2}\rightarrow0$. 
In the limit of large mass discussed above, we can seek a solution for the gap coupled equations, saying that $M_{\phi}$ and $M_{q}$ have the following behavior
\begin{eqnarray}
\label{sistemeq}
&&M^{2}_{\phi}=xm^{2}=4m^{2}+4\Psi_{0}\cr\cr
&&M_{q}=ym=m+\frac{2}{xm}\Psi_{0}+S_{0},
\end{eqnarray}
wherein $x$ and $y$ are the unknown variables that we will find later. 
This can be seen just an algebraic trick to re-write eq. (\ref{gaprc}) in a more suitable form to seek solutions. Being that, at this point, we are dealing with a non-renormalizable effective model for the low energy Yukawa Model, it is justifiable to reduce the gap equations to the bare ones, by neglecting the renormalization constants
 whose calculations
remain outside the scope of the work,
and within the usual logics of low energy effective models.

Hence, in the 
case  of large mass approximation defined earlier, the coupled gap equations  for $\Psi_0$ and $S_0$ are simplified 
\begin{eqnarray}
\frac{[b+g\phi_{0}]}{M_{\phi}^{2}}&=&\frac{1}{m}\frac{(2x-4)}{x^{2}}\rightarrow 0,\cr\cr
-\frac{1}{2}c+\frac{2f}{M^{2}_{\phi}}&=&0,
\cr\cr
\frac{4!}{\lambda}+(\frac{\Psi_{0}}{M_{\phi}^{2}}-\frac{1}{2})d+\frac{c}{M^{2}_{\phi}}&=&
\frac{24}{\lambda}\frac{(-3x^{2}+16x-64)}{x3},
\end{eqnarray}
where the second of these equations corresponds to a constraint among the model parameters and 
resulting masses:
$M_\phi^2 = 4 m^2 - \frac{\lambda}{3} \phi_0^2$. 
This equation must be compatible with the solution of 
Eq. (\ref{sistemeq}).
The other two  equations  can be written as:
\begin{eqnarray}
\label{gap12}
&&\Psi_{0}=\frac{\lambda}{24}[\frac{x^{3}}{-3x^{2}+16x-64}]i\int \frac{d^{4}p}{(2\pi)^{4}}\frac{1}{[p^{2}-M_{\phi}^{2}]}
+\frac{\lambda}{24g^{2}}[\frac{x^{3}}{-3x^{2}+16x-64}]S_{0}^{2},\cr\cr
&&S_{0}=-8C_{F}i\int\frac{d^{4}p}{(2\pi)^{4}}\frac{M_{q}}{[p^{2}-M_{q}^{2}]}.
\end{eqnarray}
Thus, after simplifications,  the following system of equations is obtained:
\begin{eqnarray}
\label{2sistemeq}
&&xm^{2}=4m^{2}+\frac{2m^{2}}{\phi_{0}^{2}}[\frac{x^{3}}{-3x^{2}+16x-64}]i\int \frac{d^{4}p}{(2\pi)^{4}}\frac{1}{[p^{2}-xm^{2}]}
+\cr\cr
&&-\frac{8}{x^{2}\phi_{0}^{4}}[\frac{x^{3}}{-3x^{2}+16x-64}][\int\frac{d^{4}p}{(2\pi)^{4}}\frac{ym}{[p^{2}-y^{2}m^{2}]}]^{2},\cr\cr
&&ym=\frac{m}{2x}(x-2)+\frac{4}{x\phi_{0}^{2}}i\int\frac{d^{4}p}{(2\pi)^{4}}\frac{ym}{[p^{2}-y^{2}m^{2}]}.
\end{eqnarray}
Observe that the masses $M_{\phi}^{2}$ and $M_{q}$ do not obey a simple scaling behavior. As we can see, we have a system of two equations and two unknown variables $(x,y)$ whose the real and positive solution for the unknown variables depends on the values of $m$ and $\phi_{0}$. Therefore with the fractions $x$ and $y$ we determine too $\Psi_{0}$ and $S_{0}$, from the system of equations
 in eq. (\ref{sistemeq}). The integrals in eq. (\ref{2sistemeq}) can be solved 
by implementing, for example, the cut-off methodology.}

\section{Effective action from external field methods}
\label{Sec12}

Above it was presented how  renormalization 
affects and refines the study of the gap equations for $\phi_{0}$, $\Psi_{0}$ and $S_{0}$. 
Now    renormalization and the gap solutions will be shown to 
affect the resulting  effective action 
with its two contributions
$\Gamma_{eff}=\Gamma_{0}+\tilde{\Gamma}$, 
wherein $\Gamma_{0}$ is the background and $\tilde{\Gamma}$ 
the fluctuations.

By assuming the auxiliary field expected value in the vacuum   are
non zero there emerges modifications in
particles and fields interactions in such non trivial background.
With the expansion in terms of the fluctuations,
 $\Psi_{R}\rightarrow \Psi_{0}+\tilde{\Psi}_{R}$, $S_{R}\rightarrow S_{0}+\tilde{S}_{R}$,
we obtain the action from eq.  (\ref{bigeqR}) in terms of two determinants:
\begin{eqnarray}
&&\exp[i\Theta]=\det\{1+\tilde{S}_{F}[-{\cal Z}_{S}^{\frac{1}{2}}{\cal Z}_{C_{F}}^{\frac{1}{2}}\tilde{S}_{R}+2{\cal Z}_{C_{F}}C_{F}^{R}(\bar{q}q)_{0}]\}\times\cr\cr
&&\det\{1-2\tilde{S}_{B}[{\cal Z}_{\Psi}^{\frac{1}{2}}{\cal Z}_{\lambda}^{\frac{1}{2}}\Psi_{R}-{\cal Z}_{\lambda}\frac{\lambda_{R}}{4!}\phi_{0}^{2}]\}^{-\frac{1}{2}}
\end{eqnarray}
where
\begin{eqnarray}
&&\tilde{S}_{F}^{-1}={\cal Z}_{q}i\slashed{\partial}-\tilde{M}_{q}(R,c.t),\quad\tilde{M}_{q}(R,c.t)={\cal Z}_{g}g_{R}\phi_{0}+\frac{b(R,c.t)}{2}+{\cal Z}_{S}^{\frac{1}{2}}{\cal Z}_{C_{F}}^{\frac{1}{2}}S_{0},\cr\cr
&&\tilde{S}_{B}^{-1}={\cal Z}_{\phi}\Box-{\tilde{M}}_{\phi}^{2}(R,c.t),\quad {\tilde{M}}_{\phi}^{2}(R,c.t)=2{\cal Z}_m m^2_R + 4{\cal Z}_\Psi^\frac{1}{2} {\cal Z}_{\lambda}^{\frac{1}{2}}
{\Psi}_{0}
\end{eqnarray}
So we have the following contribution
$i\Theta=i\Theta_{F}+i\Theta_{B}$ with the terms:
\begin{eqnarray}
&&i\Theta_{F}=\int d^{4}x\: tr\ln\{1+\tilde{S}_{F}
[-{\cal Z}_{S}^{\frac{1}{2}}{\cal Z}_{C_{F}}^{\frac{1}{2}}\tilde{S}_{R}+2{\cal Z}_{C_{F}}C_{F}^{R}(\bar{q}q)_{0}]\}\delta^{4}(x-y),\cr\cr
&&i\Theta_{B}=-\frac{1}{2}\int d^{4}x\ln\{1-2\tilde{S}_{B}
[{\cal Z}_{\Psi}^{\frac{1}{2}}{\cal Z}_{\lambda}^{\frac{1}{2}}\tilde{\Psi}_{R}+{\cal Z}_{\lambda}\frac{\lambda_{R}}{4!}\phi_{0}^{2}]\}\delta^{4}(x-y).
\end{eqnarray}
Because of the SSB and DChSB it is reasonable to perform   large fermion and scalar field
effective masses  expansions.
For the zero order derivative expansion,
 we get the subsequent leading terms for each of the determinants:
\begin{eqnarray}
\label{sectorF}
i\Theta_{F} &\simeq& 
\int d^{4}x\: tr\tilde{S}_{F}[-{\cal Z}_{S}^{\frac{1}{2}}{\cal Z}_{C_{F}}^{\frac{1}{2}}\tilde{S}_{R}+2{\cal Z}_{C_{F}}C_{F}^{R}(\bar{q}q)_{0}]\delta^{4}(x-y)+\cr\cr
&-&\frac{1}{2}\int d^{4}xtr\{\tilde{S}_{F}[-{\cal Z}_{S}^{\frac{1}{2}}{\cal Z}_{C_{F}}^{\frac{1}{2}}\tilde{S}_{R}+2{\cal Z}_{C_{F}}C_{F}^{R}(\bar{q}q)_{0}]\tilde{S}_{F}[-{\cal Z}_{S}^{\frac{1}{2}}{\cal Z}_{C_{F}}^{\frac{1}{2}}\tilde{S}_{R}+2{\cal Z}_{C_{F}}C_{F}^{R}(\bar{q}q)_{0}]\}\delta^{4}(x-y),\qquad
\end{eqnarray}
and 
\begin{eqnarray}
\label{sectorB}
i\Theta_{B} &\simeq&
\int d^{4}x\tilde{S}_{B}[{\cal Z}_{\Psi}^{\frac{1}{2}}{\cal Z}_{\lambda}^{\frac{1}{2}}\tilde{\Psi}+{\cal Z}_{\lambda}\frac{\lambda_{R}}{4!}\phi_{0}^{2}]\delta^{4}(x-y)
\cr\cr
&+&\int d^{4}x\tilde{S}_{B}
[{\cal Z}_{\Psi}^{\frac{1}{2}}{\cal Z}_{\lambda}^{\frac{1}{2}}\tilde{\Psi}_{R}+{\cal Z}_{\lambda}\frac{\lambda_{R}}{4!}\phi_{0}^{2}]\tilde{S}_{B}[{\cal Z}_{\Psi}^{\frac{1}{2}}{\cal Z}_{\lambda}^{\frac{1}{2}}\tilde{\Psi}_{R}+{\cal Z}_{\lambda}\frac{\lambda_{R}}{4!}\phi_{0}^{2}]\delta^{4}(x-y).
\end{eqnarray}

The above leading terms can be rearranged in the effective action such that one writes:
\begin{eqnarray}
\label{effAbf}
&&\Gamma_{eff}=
\int d^{4}x\{-\frac{1}{2}{\cal Z}_{\phi}\phi_{0}\Box\phi_{0}+[{\cal Z}_{m}m_{R}^{2}+\delta m(R,c.t)]\phi_{0}^{2}+[-{\cal Z}_{\lambda}\frac{\lambda_{R}}{4!}+\delta\lambda(R,c.t)]\phi_{0}^{4}
\cr\cr
&&+{\cal Z}_{q}\bar{q}_{0}i\slashed{\partial}q_{0}+[-{\cal Z}_{g}\phi_{0}-\frac{b(R,c.t)}{2}+\delta M(R,c.t)]\bar{q}_{0}q_{0}+\cr\cr
&&+[{\cal Z}_{C_{F}}C_{F}^{R}+\delta C_{F}(R,c.t)](\bar{q}_{0}q_{0})^{2}
 - \delta G \; \tilde{S}_R \tilde\Psi_R 
\},
\end{eqnarray}
wherein we can see the quantum contributions 
to the masses and coupling constants
in the  low energy or long-wavelength local limit  and a mixing-type
interaction for the two auxiliary fields for 2-fermion and 2-boson states.
The resulting interactions becomes punctual:
\begin{equation}
-i \; \delta M(R,c.t) =-2{\cal Z}_{C_{F}}C_{F}^{R}tr\tilde{S}_{F}\delta^{4}(x-y)=
8{\cal Z}_{C_{F}}C_{F}^{R}\int\frac{d^{4}p}{(2\pi)^{4}}
\frac{{\tilde{M}}_{q}(R,c.t)}{[{\cal Z}_{q}^{2}p^{2}-{\tilde{M}}_{q}^{2}(R,c.t)]},
\end{equation}
\begin{equation}
-i \; \delta C_{F}(R,c.t)=-2[{\cal Z}_{C_{F}}{C_{F}^{R}}]^{2} \; tr(\tilde{S}_{F}\tilde{S}_{F})
\delta^{4}(x-y)=-8[{\cal Z}_{C_{F}}{C_{F}^{R}}]^{2}\int\frac{d^{4}p}{(2\pi)^{4}}\frac{{\cal Z}_{q}^{2}p^{2}+{\tilde{M}}_{q}^{2}(R,c.t)}{[{\cal Z}_{q}^{2}p^{2}-{\tilde{M}}_{q}^{2}(R,c.t)]^{2}},
\end{equation}
\begin{equation}
-i \; \delta m(R,c.t)={\cal Z}_{\lambda}\frac{\lambda_{R}}{4!}\; tr 
(\tilde{S}_{B})\delta^{4}(x-y)=\frac{1}{24}{\cal Z}_{\lambda}\lambda_{R}\int \frac{d^{4}p}{(2\pi)^{4}}\frac{1}{[{\cal Z}_{\phi}p^{2}-{\tilde{M}}_{\phi}^{2}(R,c.t)]},
\end{equation}
\begin{equation}
-i \; 
\delta \lambda(R,c.t)=[{\cal Z}_{\lambda}\frac{\lambda_{R}}{4!}]^{2}
\; tr (\tilde{S}_{B}\tilde{S}_{B})
\delta^{4}(x-y)=[\frac{{\cal Z}_{\lambda}\lambda_{R}}{4!}]^{2}
\int \frac{d^{4}p}{(2\pi)^{4}}\frac{1}{[{\cal Z}_{\phi}p^{2}-{\tilde{M}}_{\phi}^{2}(R,c.t)]^2}.
\end{equation} 
\begin{eqnarray} \label{mixing1}
\delta G(R,c.t) &=&  ({\cal Z}_{S} {\cal Z}_{C_F} {\cal Z}_{\lambda} {\cal Z}_{\Psi})^\frac{1}{2}
\; tr (\tilde{S}_B \tilde{S}_F ) \delta^4 (x-y) 
\nonumber
\\
&=&
4  ({\cal Z}_{S} {\cal Z}_{C_F} {\cal Z}_{\lambda} {\cal Z}_{\Psi})^\frac{1}{2}
\int \frac{d^4 p}{(2 \pi)^4} 
\frac{1}{[{\cal Z}_{\phi}p^{2}-{\tilde{M}}_{\phi}^{2}(R,c.t)]^2}
\frac{{\tilde{M}}_{q}(R,c.t)}{[{\cal Z}_{q}^{2}p^{2}-{\tilde{M}}_{q}^{2}(R,c.t)]}.
\end{eqnarray}
 For the (long-wavelength) local limit for the second order terms 
of the determinants expansions, the fermion bilinears $(\bar{q}q$) and fields 
$\tilde{S}_R, \tilde{\Psi}_R,\phi_0$ were considered to be at the same spacetime point, being  the leading terms of  the derivative expansion \cite{Mosel}.
We are left with an effective model  for background fermions interacting with two auxiliary fields, one corresponding to a  two  $\phi$-boson quantum state, $\Psi$, and another one corresponding to a fermion-antifermion state, $S$. 
In this equations, $\delta M$ and $\delta m$ are quadratically UV-divergent
mass corrections with the same shape of the gap equations (\ref{gap12}).
They renormalize differently however being that
the parameters $b$ and $m_R^2$ must be used to the elimination of the UV divergence.
The coupling constants,  on the other hand,
$\delta C_F$ and $\delta \lambda$, 
are respectively quadratic-UV and 
log- UV divergent being eliminated by the 
${\cal Z}_{q,\phi}$ coefficients and by the renormalization prescription of the 
  $S_0$ gap equation.
The scalar field $\phi$ in the original model is responsible for the emergence of fermion effective self interactions of current-current type, 
 and higher orders. 
We can also 
extract from eq. (\ref{sectorF}) 
the free Lagrangian terms for the auxiliary fields and latter verify if they can be bound states
 quasiparticles.
 
For the field $S_R$, by considering the leading terms of 
 the derivative expansion, it  can be written as:
\begin{equation}
  - \frac{1}{4 C_F^R} {\cal Z}_{S}\tilde{S}_R^2 
-\frac{1}{2}{\cal Z}_{S}{\cal Z}_{C_{F}}tr[\tilde{S}_{F}\tilde{S}_{R}\tilde{S}_{F}
\tilde{S}_{R}]\delta^{4}(x-y)=
\frac{1}{2}\alpha(R,c.t) \partial_{\mu}\tilde{S}_{R}\partial^{\mu}\tilde{S}_{R}+ 
\frac{1}{2}\beta(R,c.t)\tilde{S}_{R}^{2} ,
\end{equation}
wherein
the field normalization and its effective mass are respectively given by:
\begin{eqnarray}
\label{2fermions}
&&\alpha(R,c.t)={\cal Z}_{S}{\cal Z}_{C_{F}}{\cal Z}_{q}^{2}
\int\frac{d^{4}p}{(2\pi)^{4}}\frac{1}{[{\cal Z}_{q}^{2}p^{2}-{\tilde{M}}_{q}^{2}(R,c.t)]^{2}}\cr\cr
&&\beta(R,c.t)=-{\cal Z}_{S}{\cal Z}_{C_{F}}\int\frac{d^{4}p}{(2\pi)^{4}}\frac{{\tilde{M}}_{q}^{2}(R,c.t)}{[{\cal Z}_{q}^{2}p^{2}-{\tilde{M}}_{q}^{2}(R,c.t)]^{2}}- \frac{1}{4 C_F^R} {\cal Z}_S.
\end{eqnarray}
Again the quadratic divergence is renormalized by a subtraction incorporated by 
a  mass counterterm.
The Yukawa-type {\it effective} interaction of this particle with an external background $(\bar{q}q)_{0}$
is given by:
\begin{equation}
2{\cal Z}_{S}^{\frac{1}{2}}{\cal Z}_{C_{F}}^{\frac{3}{2}}C_{F}^{R}\: tr[\tilde{S}_{F}
\tilde{S}_{F}]\delta^{4}(x-y)
\; 
\tilde{S}(\bar{q}q)_{0}
=
2{\cal Z}_{S}^{\frac{1}{2}}{\cal Z}_{C_{F}}^{\frac{3}{2}}C_{F}^{R}\int\frac{d^{4}p}{(2\pi)^{4}}\frac{{\cal Z}_{q}^{2}p^{2}+{\tilde{M}}_{q}^{2}(R,c.t)}{[{\cal Z}_{q}^{2}p^{2}-{\tilde{M}}_{q}^{2}(R,c.t)]^{2}}
\;
\tilde{S} \; (\bar{q}q)_{0}.
\end{equation}
These composite boson-fermion system is analogous to 
the emergence of mesons in the quark dynamics obtained from
low energy effective models
for QCD 
 \cite{Braghin2018}, although there is, in the present model,
only  one single scalar fermion-antifermion state emerging from the scalar field exchange.

In the same way,
 from eq. (\ref{sectorB}) we note  the emergence of dynamics of a 
composite field $\tilde{\Psi}$, eventually corresponding to a two-boson state.
The   leading effective Lagrangian terms  obtained from the derivative expansion
 can be written as:
\begin{eqnarray}
[\frac{4!}{\lambda_{R}}{\cal Z}_{\Psi}-\frac{d(R,c.t)}{2}]\tilde{\Psi}_{R}^{2} + 
{\cal Z}_{\Psi}{\cal Z}_{\lambda}\tilde{S}_{B}\tilde{\Psi}_{R}\tilde{S}_{B}
\tilde{\Psi}_{R}\delta^{4}(x-y)=\frac{\epsilon(R,c.t)}{2} 
\tilde{\Psi}_{R}\Box\tilde{\Psi}_{R}+ \frac{\sigma(R,c.t)}{2}\tilde{\Psi}^{2}_{R}
\end{eqnarray}

wherein the field normalization and mass can be expressed as:
\begin{eqnarray}
\label{2bosons}
&&\epsilon(R,c.t)= 2 
{\cal Z}_{\Psi}{\cal Z}_{\lambda}{\cal Z}_{\phi}\int\frac{d^{4}p}{(2\pi)^{4}}\frac{1}{[{\cal Z}_{\phi}p^{2}-{\tilde{M}}_{\phi}^{2}(R,c.t)]^{2}[{\cal Z}_{\phi}p^{2}+{\tilde{M}}_{\phi}^{2}(R,c.t)]}\cr\cr
&&
\sigma(R,c.t)=2
{\cal Z}_{\Psi}{\cal Z}_{\lambda}\int\frac{d^{4}p}{(2\pi)^{4}}\frac{{\tilde{M}}_{\phi}^{2}}{[{\cal Z}_{\phi}p^{2}-{\tilde{M}}_{\phi}^{2}(R,c.t)]^{2}[{\cal Z}_{\phi}p^{2}+{\tilde{M}}_{\phi}^{2}(R,c.t)]}+\cr\cr
&&+2[\frac{4!}{\lambda_{R}}{\cal Z}_{\Psi}-\frac{d(R,c.t)}{2}] .
\end{eqnarray}

The  effective interaction of this composite field  with an external background $\phi_{0}^{2}$
in the long wavelength  local limit
can be written  as:
\begin{equation}
\left( 2{\cal Z}_{\Psi}^{\frac{1}{2}}{\cal Z}_{\lambda}^{\frac{1}{2}}{\cal Z}_{\lambda}\frac{\lambda_{R}}{4!}\tilde{S}_{B}\tilde{S}_{B}\delta^{4}(x-y) 
\right) \; 
\tilde{\Psi}_{R}\phi_{0}^{2} \; 
=
\; 
\left( 
\frac{1}{12}{\cal Z}_{\Psi}^{\frac{1}{2}}{\cal Z}_{\lambda}^{\frac{3}{2}}\lambda_{R}
\int \frac{d^{4}p}{(2\pi)^{4}}\frac{1}{[{\cal Z}_{\phi}p^{2}-{\tilde{M}}_{\phi}^{2}(R,c.t)]^{2}}
\right)
\; \tilde{\Psi}_{R}\phi_{0}^{2}
\end{equation}
where an effective coupling constant was naturally resolved.

From the above equations, it is useful to resolve effective interactions between 
the remaining fields. 
Consider the following quantities:
$$\Upsilon_{F}(R,c.t)=2{\cal Z}_{S}^{\frac{1}{2}}{\cal Z}_{C_{F}}^{\frac{3}{2}}\int\frac{d^{4}p}{(2\pi)^{4}}\frac{{\cal Z}_{q}^{2}p^{2}+{\tilde{M}}_{q}^{2}(R,c.t)}{[{\cal Z}_{q}^{2}p^{2}-{\tilde{M}}_{q}^{2}(R,c.t)]^{2}},$$
and 
$$\Upsilon_{B}(R,c.t)=\frac{1}{12}{\cal Z}_{\Psi}^{\frac{1}{2}}{\cal Z}_{\lambda}^{\frac{1}{2}}{\cal Z}_{\lambda}
\int \frac{d^{4}p}{(2\pi)^{4}}\frac{1}{[{\cal Z}_{\phi}p^{2}-{\tilde{M}}_{\phi}^{2}(R,c.t)]^{2}}.$$
An effective model for the YuM can be written as
\begin{eqnarray} \label{S-auxfields}
&&\tilde{Z}=\int D\tilde{S}_{R}D\tilde{\Psi}_{R}\exp\{i\int d^{4}x(-i)[\tilde{S}_{R}[-\alpha(R,c.t)\Box+\beta(R,c.t)]\tilde{S}_{R}+\Upsilon_{F}(R,c.t)C_{F}^{R}(\bar{q}q)_{0}\tilde{S}_{R}+\cr\cr
&&+\tilde{\Psi}_{R}[\epsilon(R,c.t)\Box+\sigma(R,c.t)]\tilde{\Psi}_{R}+\Upsilon_{B}(R,c.t)\lambda_{R}\phi_{0}^{2}\tilde{\Psi}_{R}]
 - \delta G \; \tilde{S}_R \tilde\Psi_R \},
\end{eqnarray}
in which we see  the kinetic term for the free composite fields and their
 interactions with the external background condensates similarly to other models,
for earlier reviews see \cite{NJL1,Klev,Ebert}.
If we continue the expansion we would have further interactions between
 these composite fields and derivative couplings as well.

The mixing interaction 
 $S_R$ and $\Psi_R$
suggests one can rotate the system of states $S_R$ and $\Psi_R$ to diagonalize the corresponding masses.
Let us define a mixing angle $\theta_{22}$ that allow to rotate the auxiliary fields
of 2-fermion states to 2-boson states:
\begin{eqnarray}
\tilde{S}_{R} &=& \cos (\theta_{22}) \tilde{S}_{mass,R} + \sin (\theta_{22}) 
\tilde{\Psi}_{mass,R},
\\
\tilde{\Psi}_{R } &=&   \cos (\theta_{22}) \tilde{\Psi}_{mass,R}
 - \sin (\theta_{22}) \tilde{S}_{mass,R},
\end{eqnarray}
where $\tilde{S}_{R,mass}$ and $\tilde{\Psi}_{R,mass}$ are mass eigenstates.
With this rotation the following effective  action is obtained by omitting the 
indices $_{mass}$:
\begin{eqnarray} \label{S-auxfields-2}
S_{eff}&=&
(-i)[\tilde{S}_{R}[-\alpha(R,c.t)  \cos^2 (\theta_{22} )\Box
+ \cos^2 (\theta_{22} )\beta(R,c.t)
+ \delta G \sin (2 \theta_{22}) ]\tilde{S}_{R}
\nonumber
\\
&+& 
\Upsilon_{F}(R,c.t)C_{F}^{R}(\bar{q}q)_{0}
 [\cos (\theta_{22}) \tilde{S}_{mass,R} + \sin (\theta_{22}) 
\tilde{\Psi}_{mass,R}]
\nonumber
\\
&+&\Upsilon_{B}(R,c.t)\lambda_{R}\phi_{0}^{2} [  \cos (\theta_{22}) \tilde{\Psi}_{mass,R}
 - \sin (\theta_{22}) \tilde{S}_{mass,R}]
\nonumber
\\
&+&\tilde{\Psi}_{R}[\epsilon(R,c.t) \cos^2 (\theta_{22} )\Box
+
 \cos^2(\theta_{22})\sigma(R,c.t)
- \delta G \sin (2 \theta_{22}) ]\tilde{\Psi}_{R}]
\nonumber
\\
&+&\sin (2 \theta_{22})
\tilde{\Psi}_{R}[\epsilon(R,c.t) \Box
+\sigma(R,c.t)]\tilde{S}_{R}]
\nonumber
\\
&+&
 \sin (2 \theta_{22}) 
\tilde{S}_{R}[-\alpha(R,c.t) \Box
+ \beta(R,c.t)]\tilde{\Psi}_{R}
\nonumber
\\
&-& 
 \delta G \; \cos (2 \theta_{22})
\tilde{S}_{mass,R} 
\tilde{\Psi}_{mass,R}   
\},
\end{eqnarray}
With two  
 conditions of final mass eigenstates and normalization 
\begin{eqnarray}
&&\tan (2 \theta_{22})= 
\frac{\delta G}{
\sigma(R,c.t)
+ \beta(R,c.t)},\cr\cr
&&\epsilon (R,c.t) - \alpha (R,c.t)=0.
\end{eqnarray}

At this point it becomes
interesting to define the bound state conditions for both 
the composite scalar field $\tilde{\Psi}_R$ and for the 
composite fermion-antifermion state $\tilde{S}_R$.
They can be simply identified to the condition of 
a pole for real positive masses
\begin{eqnarray}
m_{\tilde{S}}^{2}(R,c.t)&=&-\frac{\beta(R,c.t)}{\alpha(R,c.t)}+ 2 \delta G \tan(\theta_{22}),
\;\;\;\;\;\;
m_{\tilde{\Psi}}^{2}(R,c.t)=\frac{\sigma(R,c.t)}{\epsilon(R,c.t)}+ 2 \delta G \tan(\theta_{22}).
\end{eqnarray}
 They can be written, by considering eq. (\ref{2fermions}) and eq. (\ref{2bosons}) as:
\begin{eqnarray}
\label{mboundR}
&&{\cal Z}_{q}^{2}m^{2}_{\tilde{S}}(R,c.t)=\frac{{\cal Z}_{\Psi}}{{\cal Z}_{C_{F}}}\tilde{M}^{2}_{q}(R,c.t)+\frac{1}{4{\cal Z}_{C_{F}}C^{R}_{F}}\frac{1}{\int\frac{d^{4}p}{(2\pi)^{4}}\frac{1}{[{\cal Z}_{q}^{2}p^{2}-{\tilde{M}}_{q}^{2}(R,c.t)]^{2}}}+  2 \delta G \tan(\theta_{22}),\cr\cr
&&{\cal Z}_{\phi}m_{\tilde{\Psi}}^{2}(R,c.t)=\tilde{M}^{2}_{\phi}(R,c.t)+[\frac{4!}{{\cal Z}_{\lambda}\lambda_{R}}-\frac{4\phi_{0}^{2}(R,c.t)}{M^{2}_{\phi}(R,c.t)}]\times\cr\cr
&&\times\frac{1}{\int\frac{d^{4}p}{(2\pi)^{4}}\frac{1}{[{\cal Z}_{\phi}p^{2}-{\tilde{M}}_{\phi}^{2}(R,c.t)]^{2}[{\cal Z}_{\phi}p^{2}+{\tilde{M}}_{\phi}^{2}(R,c.t)]}}+  2 \delta G \tan(\theta_{22}).
\end{eqnarray}

\subsection{Mass and existence of the composite particles}

From the pole condition for the composite fields in eq. (\ref{mboundR}), the bound state conditions for the composite particle from the scalar field $\tilde{\Psi}$ can be write in the following form
\begin{eqnarray}
\label{bond1}
&&{\cal Z}_{\phi}m_{\tilde{\Psi}_{R}}^{2}(R,c.t)=\tilde{M}^{2}_{\phi}(R,c.t)+\frac{4!}{{\cal Z}_{\lambda}\lambda_{R}}[1-\frac{(\frac{{\cal Z}_{m}}{{\cal Z}_{\lambda}}12m^{2}_{R}-{\cal Z}_{\Psi}^{\frac{1}{2}}{\cal Z}_{\lambda}^{\frac{1}{2}}48\Psi_{0}(R,c.t))}{12M^{2}_{\phi}(R,c.t)}]\times\cr\cr
&&\times\frac{1}{\int\frac{d^{4}p}{(2\pi)^{4}}\frac{1}{[{\cal Z}_{\phi}p^{2}-{\tilde{M}}_{\phi}^{2}(R,c.t)]^{2}[{\cal Z}_{\phi}p^{2}+{\tilde{M}}_{\phi}^{2}(R,c.t)]}}+ 2 \delta G \tan(\theta_{22}),
\end{eqnarray}
in which we consider the constant background $\phi_{0}$ with the quantum contributions, seen in eq. (\ref{gapr1}). As we know $\lambda_{R}>0$, so we have a composite scalar field particle due to the existence of real pole for the two point function when
\begin{eqnarray}
\label{ine1}
&&\tilde{M}^{2}_{\phi}(R,c.t)\geq\frac{4!}{{\cal Z}_{\lambda}\lambda_{R}}[1-\frac{(\frac{{\cal Z}_{m}}{{\cal Z}_{\lambda}}12m^{2}_{R}-{\cal Z}_{\Psi}^{\frac{1}{2}}{\cal Z}_{\lambda}^{\frac{1}{2}}48\Psi_{0}(R,c.t))}{12M^{2}_{\phi}(R,c.t)}]\times\cr\cr
&&\times\frac{1}{\int\frac{d^{4}p}{(2\pi)^{4}}\frac{1}{[{\cal Z}_{\phi}p^{2}-{\tilde{M}}_{\phi}^{2}(R,c.t)]^{2}[{\cal Z}_{\phi}p^{2}+{\tilde{M}}_{\phi}^{2}(R,c.t)]}}+ 2 \delta G \tan(\theta_{22})
\end{eqnarray}

In  another hand $C^{R}_{F}=-\frac{1}{2}\frac{g^{2}_{R}}{M^{2}_{\phi}(R,c.t)}$ and ${\cal Z}_{g}g_{R}\phi_{0}(R,c.t)={\cal Z}_{m}^{\frac{1}{2}}m_{R}$, so the bound state conditions for composite particle from the fermion-antifermion field $\tilde{S}_R$ can be write in the following form
\begin{equation}
\label{bond2}
{\cal Z}_{q}^{2}m^{2}_{\tilde{S}_{R}}(R,c.t)=\frac{{\cal
Z}_{\Psi}}{{\cal Z}_{C_{F}}}\tilde{M}^{2}_{q}(R,c.t)-\frac{1}{2}\frac{M^{2}_{\phi}(R,c.t)}{{\cal Z}_{m}m_{R}^{2}}\phi^{2}_{0}(R,c.t)\frac{1}{\int\frac{d^{4}p}{(2\pi)^{4}}\frac{1}{[{\cal Z}_{q}^{2}p^{2}-{\tilde{M}}_{q}^{2}(R,c.t)]^{2}}}+ 2 \delta G \tan(\theta_{22}).
\end{equation}
So in the weak regime for the couplings ($g_{R}$, $\lambda_{R}$), $C_{F}\rightarrow0$ and $\phi_{0}^{2}\rightarrow\infty$ and so we do not have a composite fermion-antifermion due to the existence of a imaginary pole for the two point Green function. Contrarily, in the strong regime for the couplings ($g_{R}$, $\lambda_{R}$) we have that
\begin{eqnarray}
&&{\cal Z}_{q}^{2}m^{2}_{\tilde{S}_{R}}(R,c.t)=\frac{
{\cal Z}_{\Psi}}{ {\cal Z}_{C_{F}}}[{\cal Z}_{g}g_{R}\phi_{0}+\frac{1}{2}\frac{{\cal Z}_{g}g_{R}}{M_{\phi}^{2}}(2{\cal Z}_{m}m_{R}^{2}\phi_{0}-{\cal Z}_{\lambda}\frac{\lambda_{R}}{3!}\phi_{0}^{3}+4{\cal Z}_{\Psi}^{\frac{1}{2}}{\cal Z}_{\lambda}^{\frac{1}{2}}\phi_{0}\Psi_{0})+{\cal Z}_{S}^{\frac{1}{2}}{\cal Z}_{C_{F}}^{\frac{1}{2}}S_{0}]^{2}+\cr\cr
&&-\frac{1}{2}\frac{M^{2}_{\phi}(R,c.t)}{{\cal Z}_{m}m^{2}_{R}}\phi^{2}_{0}(R,c.t)\frac{1}{\int\frac{d^{4}p}{(2\pi)^{4}}\frac{1}{[{\cal Z}_{q}^{2}p^{2}-{\tilde{M}}_{q}^{2}]^{2}(R,c.t)}}+  2 \delta G \tan(\theta_{22}).
\end{eqnarray}
After simplifications we arrived in the inequality for the existence of composite fermion-antifermion state
\begin{eqnarray}
\label{orderparameter}
&&\frac{1}{2}\frac{M^{2}_{\phi}(R,c.t)}{{\cal Z}_{m}m^{2}_{R}}\frac{1}{\int\frac{d^{4}p}{(2\pi)^{4}}\frac{1}{[{\cal Z}_{q}^{2}p^{2}-{\tilde{M}}_{q}^{2}(R,c.t)]^{2}}}\phi^{2}_{0}(R,c.t)\leq\frac{{\cal Z}_{\Psi}}{{\cal Z}_{g}^{2}}[{\cal Z}_{m}^{\frac{1}{2}}m_{R}+\cr\cr
&&+\frac{1}{2}\frac{{\cal Z}_{m}^{\frac{1}{2}}m_{R}}{M_{\phi}^{2}}({\cal Z}_{m}m_{R}^{2}+4{\cal Z}_{\Psi}^{\frac{1}{2}}{\cal Z}_{\lambda}^{\frac{1}{2}}\Psi_{0})+{\cal Z}_{S}^{\frac{1}{2}}{\cal Z}_{C_{F}}^{\frac{1}{2}}S_{0}]^{2}+  2 \delta G \tan(\theta_{22}),
\end{eqnarray}
or we can write the previous inequality in terms of $\lambda_{R}$, i.e.
\begin{eqnarray}
\label{restrictlambda}
&&{\cal Z}_{\lambda}\lambda_{R}\geq\frac{1}{\int\frac{d^{4}p}{(2\pi)^{4}}\frac{1}{[{\cal Z}_{q}^{2}p^{2}-{\tilde{M}}_{q}^{2}(R,c.t)]^{2}}}\frac{1}{2{\cal Z}_{\Psi}m^{2}_{R}}\times\cr\cr
&&\times\frac{{\cal Z}_{g}^{2}M^{2}_{\phi}(R,c.t)[\frac{{\cal Z}_{m}}{{\cal Z}_{\lambda}}12m^{2}_{R}-{\cal Z}_{\Psi}^{\frac{1}{2}}{\cal Z}_{\lambda}^{\frac{1}{2}}48\Psi_{0}(R,c.t)]}{{\cal Z}_{m}[{\cal Z}_{m}^{\frac{1}{2}}m_{R}+\frac{1}{2}\frac{{\cal Z}_{m}^{\frac{1}{2}}m_{R}}{M_{\phi}^{2}}({\cal Z}_{m}m_{R}^{2}+4{\cal Z}_{\Psi}^{\frac{1}{2}}{\cal Z}_{\lambda}^{\frac{1}{2}}\Psi_{0})+{\cal Z}_{S}^{\frac{1}{2}}{\cal Z}_{C_{F}}^{\frac{1}{2}}S_{0}]^{2}+{\cal Z}_{g}^{2} 2 \delta G \tan(\theta_{22})}.
\end{eqnarray}
 We can conclude too that the existence or not of composite fermion-antifermion state are dictated by the value of $\phi_{0}^{2}$ (SSB). As we restrict to cases in which $m_{R}$ is very large, $M^{2}_{\phi}=x{\cal Z}_{m}m^{2}_{R}$ and $M_{q}=y{\cal Z}^{\frac{1}{2}}_{m}m_{R}$, so with the solution in eq. (\ref{sistemeq}) for ($x$, $y$, $S_{0}$, $\Psi_{0}$) we can solve the above inequalities in eq. (\ref{ine1}) and eq. (\ref{restrictlambda}) by the cut-off methodology (${\cal Z}_{i} \rightarrow 1$).

\subsection{Effective action for the original model}

However the auxiliary fields might not be quasiparticles of the system, in which case 
a current expansion can be performed based in the
contributions of the two-fermion or two-boson states.
By eliminating the auxiliary fields at the  level discussed above 
by considering the only quadratic terms
 it can be written that:
\begin{eqnarray} \label{int-auxfield}
&&\tilde{Z}=\det[-\alpha(R,c.t)\Box+\beta(R,c.t)]^{-\frac{1}{2}}\det[\epsilon(R,c.t)\Box+\lambda(R,c.t)]^{-\frac{1}{2}}\times\cr\cr
&&\times\exp\{i\int d^{4}xd^{4}y[-\frac{1}{4}j_{F}Vj_{F}-\frac{1}{4}j_{B}Wj_{B}]\}.
\end{eqnarray}

Now we define following quantities:
\begin{eqnarray}
&&\tilde{S}_{R}\rightarrow \bar{S}_{R}-\frac{1}{2}\int d^{4}yV(x,y)j_{F}\cr\cr
&&j_{F}=\Upsilon_{F}(R,c.t) C_{F}^{R}(\bar{q}q)_{0}\cr\cr
&&V^{-1}(x,y)=[-\alpha(R,c.t)\Box+\beta(R,c.t)]\delta^{4}(x-y),
\end{eqnarray}
\begin{eqnarray}
&&\tilde{\Psi}_{R}\rightarrow \bar{\Psi}_{R}-\frac{1}{2}\int d^{4}yW(x,y)j_{B}\cr\cr
&&j_{B}=\Upsilon_{B}(R,c.t)\lambda_{R}\phi_{0}^{2}\cr\cr
&&W^{-1}=[\epsilon(R,c.t)\Box+\sigma(R,c.t)]\delta^{4}(x-y),
\end{eqnarray}
So that the   vertices in eq. (\ref{int-auxfield}) can be written respectively as:
\begin{eqnarray}
&&-\frac{1}{4}j_{F}Vj_{F}=\delta C_{F}^{V}(R,c.t)(\bar{q}q)_{0}^{2},\cr\cr
&&\mbox{where} \quad \delta C_{F}^{V}(R,c.t)=-[\Upsilon_{F}(R,c.t)C_{F}^{R}]^{2}\int\frac{d^{4}p}{(2\pi)^{4}}\frac{\exp[ip(x-y)]}{[\alpha(R,c.t) p^{2}+\beta(R,c.t)]},\cr\cr
&&-\frac{1}{4}j_{B}Wj_{B}=\delta\lambda^{W}(R,c.t)\phi_{0}^{4},\cr\cr
&&\mbox{where} \quad
\delta\lambda^{W}(R,c.t)=-[\Upsilon_{B}(R,c.t)\lambda_{R}]^{2}{\int\frac{d^{4}p}{(2\pi)^{4}}\frac{\exp[ip(x-y)]}{[-\epsilon(R,c.t) p^{2}+\sigma(R,c.t)]}}.
\end{eqnarray}

Therefore the effective action for the YuM can be written as
{\small
\begin{eqnarray}
\label{quantumeff}
&&\Gamma_{eff}
=\int d^{4}x\{-\frac{1}{2}{\cal Z}_{\phi}\phi_{0}\Box\phi_{0}+[{\cal Z}_{m}m^{2}+\delta m(R,c.t)]\phi_{0}^{2}+[-{\cal Z}_{\lambda}\frac{\lambda_{R}}{4!}+\delta\lambda(R,c.t)-i\delta\lambda^{W}(R,c.t)]\phi_{0}^{4}
\cr\cr
&&+{\cal Z}_{q}\bar{q}_{0}i\slashed{\partial}q_{0}+[-{\cal Z}_{g}g\phi_{0}-\frac{b(R,c.t)}{2}+\delta M(R,c.t)]\bar{q}_{0}q_{0}+[{\cal Z}_{C_{F}}C_{F}^{R}+\delta C_{F}(R,c.t)-i\delta C_{F}^{V}(R,c.t)](\bar{q}_{0}q_{0})^{2}\}
\cr\cr
&&-\frac{1}{2}tr\ln[-\alpha(R,c.t)\Box+\beta(R,c.t)]-\frac{1}{2}tr\ln[\epsilon(R,c.t)\Box+\sigma(R,c.t)].
\end{eqnarray}}
In this calculation the role of the  auxiliary fields
is encoded in the non linear behavior and dependencies of the 
resulting corrections for the  masses and vertices
of
the original and effective parameters defined along the work  on the ground state values of 
these auxiliary fields:
[$\delta m, b, \delta M$ and 
$\delta \lambda, \delta\lambda^{W},  C_F^R,  \delta C_F$ and  $\delta C_{F}^{V}$],
besides further contributions from the integration over the composite auxiliary fields.

We finished this section by adding  a brief comment on the previous analysis.
 As we can see we achieved a way of analyzing  how the counter-terms 
and the spontaneous symmetry breaking contribute to 
 the chiral symmetry breaking and dynamical {mass generation mechanism}. 
This is  seen not only in the influence of these ingredients in the gap equations
above
 but also in the investigation of its effects in the construction of an effective action by external field methods
as just noticed. 
The analysis presented here is based in the assumption that there are values (regions) of the coupling constants ($g_{R}$, $\lambda_{R}$) that permit  solutions to the coupled gap  equations, eq. (\ref{gapr1}) and eq. (\ref{gapr2}).
In the limit of ${\cal Z}_{i}\rightarrow1$ we recover the usual gap  equation
 with the ultraviolet divergences  corresponding to 
the emergence of a (non renormalizable) low energy   effective model.
For the resulting  effective action we can see
 four types of ultraviolet divergences in the fermion sector due to the fourth order expansion in 
large fermion effective mass,
and three
 types of ultraviolet divergences in the boson sector.
So that  we have 
 seven types of infinity and 
seven counter-terms ($\delta{\cal Z}_{q}$, $\delta{\cal Z}_{g}$, $\delta{\cal Z}_{\phi}$, $\delta{\cal Z}_{m},\delta{\cal Z}_{\lambda}$, $\delta{\cal Z}_{\Psi}$, $\delta{\cal Z}_{S}$).
In the effective action, eq. (\ref{quantumeff}),
 we have five renormalization conditions, three in the scalar sector associated with the on shell behavior not only of the propagator (residue equal to 1 and pole in physical mass) but also the vertex, and two in the fermion sector again associated with behavior of the propagator. Finally with the two conditions from the gap equations, we determine all the counter-terms. The renormalization of the resulting effective model is a result of the original model (YuM), which is renormalizable, but as we have seen, the gap equations already correspond to the non-renormalizable model (Fermi), requiring more counter-terms to adjust the theory.

\section{Outcomes and final comments}
\label{Sec3}

 By considering the Yukawa model, throughout this work,
 different ways of dealing with field   interactions in terms of equivalent linear 
 quadratic or  quartic  structures
were investigated, wherein the renormalization procedure plays an important role not only in the gap equations but also in  the effective action due to quantum contributions, eliminating the infinites and adjusting the coupling parameters, masses and gaps to physical results. 
A modification in the Hubbard-Stratonovich  auxiliary  field  (HSAF) identity
\cite{FLB-epjp}, seen in eq. (\ref{Hubident}),
that account for SSB order parameter contribution was  also implemented.
In the limit of $\phi_{0}\rightarrow 0$ we have an usual Hubbard-Stratonovich identity with no spontaneous symmetry breaking of $Z_{2}$ symmetry. 
The resulting  functional generator  was then written 
in a better form {by} implementing the expansion in terms of currents. 
This change in the HSAF  identity  affects all the following
analysis. We have a  generator  functional 
 in eq. (\ref{bigeq}) with a residual fermion current-current effective interaction
 and from it we compute the effective potential with the coupled  gap equations in {eq. (\ref{gapr2})}. 
If  we restrict, as a physical approximation, to a large mass approximation for  $m$, so that the massive {mediator} particle associated with the $\phi$ should
be in some sense confined, the corresponding coupled gap equations are simplified and can be re-written in a form similar to the one known in the literature (${\cal Z}_{i} \rightarrow 1$) {. As a consequence
we can explore solutions for the coupled  gap equations in a particular regime of 
coupling constants ($g$, $\lambda$), seen in (\ref{sistemeq}) and eq. (\ref{2sistemeq}) respectively.} 
In the literature not only the large mass approximation are investigated,
 where two coupled gap equations were analyzed
\cite{Paulo-Braghin,PRD-2013}, but also the strong Yukawa coupling was  explored \cite{strongY}.

As a second outcome 
  the effective action of the model was calculated 
by {the} background external field method, 
seen in eq. (\ref{effAbf}) with eq. (\ref{S-auxfields}), in which we are left
 with an effective model  for background fields interacting with two auxiliary fields, 
one may correspond to {a} two-boson quantum states and 
another one  to a fermion-antifermion state. Bound state conditions for these two composite states  were established in eq. (\ref{bond1}) and eq. (\ref{bond2}), wherein we see in eq. (\ref{orderparameter}) that the existence or not of composite fermion-antifermion state are dictated by the value of SSB order parameter $\phi_{0}^{2}$.
{Finally  from the YuM  effective action, for example by means of eq
 (\ref{effAbf})
or eq. (\ref{quantumeff}), 
we can see  the appearance of seven
 types of (UV) infinities and seven counter-terms
 ($\delta{\cal Z}_{q}$, $\delta{\cal Z}_{g}$, $\delta{\cal Z}_{\phi}$, $\delta{\cal Z}_{m},\delta{\cal Z}_{\lambda}$, 
$\delta{\cal Z}_{\Psi}$, $\delta{\cal Z}_{S}$) adjusted by the physical conditions.
The physical  masses of the two boson composite state $\tilde{\Psi}$ and fermion-antifermion composite state $\tilde{S}$, whenever they are formed,
were given in Eqs. (\ref{bond1}, \ref{ine1}) and Eqs. (\ref{bond2}, \ref{restrictlambda}) respectively, corresponding to the poles of their propagators.
They can only be obtained by considering the (ground state) solutions of the gap equations for all the fields involved being that the gap equations turn out to be coupled. This highly non linear set of equations cannot be easily diagonalized. 
The resulting off-diagonal terms,  associated with the boson-boson and fermion-antifermion scalar fields ($S,\Psi$)
interactions/mixings, 
were discussed in two levels.
 Firstly, the condition for minima of the effective potential
that yield  the gap equations
can be done with the Hessian characteristic equation.
Secondly, 
 a mixing interaction in the effective action of the type $\delta G {S}_R\Psi_R$ required a rotation 
of eigenstates by
defining  a mixing angle $\theta_{22}$ to 
find the mass eigenstates of the 2-boson and 2-fermion composite states. 
Note that the corresponding mixing mechanism is very different from 
the quark mixing in the Standard Model \cite{Predazzi}.
This mixing emerges  from the above one loop calculation,  one
internal  fermion line and one internal boson
 line,  similarly to  vacuum polarization  worked out in \cite{mesonmixing}
for neutral  mesons mixing mechanism that  is 
usually addressed in low energies effective models for QCD
by means of  other mechanisms
\cite{NJL1,Klev,Ebert,coimbra}.
Despite we present the phenomenon in the work, a further investigation is needed.
Although the initial toy model considered here is  pertubatively renormalizable, the resulting current-current  fermion interaction is not.  This is easily seen in the fact that the coupling constants $g$ and $\lambda$ are dimensionless and the Fermi constant $C_{F}$ has the dimension of squared length.
 So the investigation of how the renormalization affects the gap equations and the
resulting  effective action analysis is quite important to understand.} As far as the authors know, the renormalization technique for the gaps in various models is not properly explored  in the literature. A next step of investigation is the relationship between the auxiliary field formalism and the gap equations in a thermal environment  to
articulate eventual symmetries restorations. This matter will be investigated and requires elaborations.

\section*{Acknowledgements}
A.A.N. thanks National Post-Doctoral Program (PNPD/UFG) for support in his brief stay at UFG.
F.L.B.  is member of
INCT-FNA,  Proc. 464898/2014-5
and  he acknowledges partial support from 
CNPq-312072/2018-0 
and  
CNPq-421480/2018-1.

\end{document}